\documentclass[PAPER, UKenglish,titleextraheight=1em,copyrightoffset=5em,cernpreprint, texmf, orcidlogo]{atlasdoc}
 
\usepackage[biblatex=true,backend=biber,subfigure]{atlaspackage}
\usepackage{atlasbiblatex}
 
\usepackage{atlasphysics}
\usepackage{atlasprocess}

\graphicspath{{logos/}{figures/}}
 
\usepackage{ANA-STDM-2018-05-PAPER-defs}

\AtlasTitle{A precise measurement of the $Z$-boson double-differential
transverse momentum and rapidity distributions in the full phase space
of the decay leptons with the ATLAS experiment at~$\sqrt s~=~8$~TeV}

\nolinenumbers
 
\AtlasAbstract{
This paper presents for the first time a precise measurement of the production properties of the $Z$~boson in the full phase space of the decay leptons. This is in contrast to the many previous precise unfolded measurements performed in the fiducial phase space of the decay leptons. The measurement is obtained from proton--proton collision data collected by the ATLAS experiment in 2012 at $\sqrt{s} = 8$ TeV at the LHC and corresponding to an integrated luminosity of 20.2~fb$^{-1}$. The results, based on a total of 15.3 million $Z$-boson decays to electron and muon pairs, extend and improve a previous measurement of the full set of angular coefficients describing $Z$-boson decay. The double-differential cross-section distributions in $Z$-boson transverse momentum~\pT\ and rapidity~\yZ\ are measured in the pole region, defined as $80 < \mll < 100$ GeV, over the range $|y| < 3.6$. The total uncertainty of the normalised cross-section measurements in the peak region of the \pT~distribution is dominated by statistical uncertainties over the full range and increases as a function of rapidity from $0.5-1.0\%$ for $|y| < 2.0$ to $2-7\%$ at higher rapidities. The results for the rapidity-dependent transverse momentum distributions are compared to state-of-the-art QCD predictions, which combine in the best cases approximate N$^4$LL resummation with N$^3$LO fixed-order perturbative calculations. The differential rapidity distributions integrated over~\pT\ are even more precise, with accuracies from~$0.2-0.3\%$ for~$|y| < 2.0$ to~$0.4-0.9\%$ at higher rapidities, and are compared to fixed-order QCD predictions using the most recent parton distribution functions. The agreement between data and predictions is quite good in most cases.
}

\author{The ATLAS Collaboration}
 
\AtlasRefCode{ATLAS-STDM-2018-05}

\PreprintIdNumber{CERN-EP-2023-171}
 
\AtlasJournal{EPJC}
\AtlasJournalRef{Eur.\ Phys.\ J.\ C 84 (2024) 315}
\AtlasDOI{10.1140/epjc/s10052-024-12438-w}

\hypersetup{pdftitle={ATLAS document},pdfauthor={The ATLAS Collaboration}}

\begin{document}
 
\maketitle

\section{Introduction and motivation}
\label{sec:Intro}

The production of $Z$~bosons and their decay to lepton pairs at the Large Hadron Collider~(LHC) through the Drell--Yan mechanism~\cite{DrellYan70} has been the topic of very fruitful and detailed studies in the LHC~experiments~\cite{ATLAS8TeV, STDM-2018-14,CMS-SMP-13-013,CMS13TeV,ATLASAi,CMS-SMP-13-010,STDM-2016-04}. The precision of the measurements has motivated over the past decade impressive theoretical developments, mostly in the area of higher-accuracy quantum chromodynamic~(QCD)~predictions, but also in  the area of parton distribution functions~(PDFs). The sub-percent precision achieved for the absolute and normalised fiducial cross-section measurements of the $Z$-boson transverse momentum,~\pTZ, as a function of its rapidity,~\absyZ, at the $Z$~pole strongly constrains state-of-the-art theoretical calculations. In the best cases, these calculations combine approximate next-to-next-to-next-to-next-to-leading logarithm~(N$^4$LL) perturbative resummation at low~\pTZ\ with O($\alpha_s^3$)~(N$^3$LO) fixed-order perturbative calculations at high~\pTZ, where $\alpha_s$~denotes the strong coupling constant. Similarly, the precision obtained for the fiducial cross-section measurements of~\absyZ\ provides strong constraints on the parton distribution functions.
 
This paper presents for the first time a double-differential measurement in~(\pTZ,\absyZ) of absolute and normalised cross-sections at the $Z$~pole within the full phase space of the decay leptons. The measurement uses the full coverage of the ATLAS detector to combine 6.2~million electron and 7.8~million muon pairs from Z-boson decays in the central region (\eeCC~and \mmCC~channels), complemented by 1.3~million electron pairs with one electron in the forward region of the detector (\eeCF~channel). Such a measurement is possible using the methodology already developed and published for the extraction of the \Zboson\-boson polarisation~\cite{ATLASAi}. It relies on the decomposition of the lepton angular $\cos\theta$ and $\phi$ distributions in the Collins-Soper frame~\cite{Collins:1977prd} into nine spherical harmonic polynomials,~$P_i$, multiplied by angular coefficients,~$A_i$~\cite{Mirkes:1992hu, arXiv9406381, Mirkes94PhysRevD, Mirkes94prep}. For pure $Z$-boson production, the full five-dimensional differential cross section describing the kinematics of the two Born-level leptons can be written as:
\begin{align}
\label{eq:master2}
\frac{ \mathrm{d}\sigma}{\mathrm{d}\pTZ~\mathrm{d}\yZ~\mathrm{d}m~\mathrm{d}\cos\theta~\mathrm{d}\phi} =&
\frac{3}{ 16 \pi} \frac{\mathrm{d}\sigma ^{ U+ L }}{\mathrm{d}\pTZ~\mathrm{d}\yZ~\mathrm{d}m}  \nonumber \\
\bigg\{&(1 + \cos^2\theta) + \frac{1}{2}\ A_0 (1 - 3 \cos^2\theta) + A_1 \ \sin 2 \theta\ \cos\phi \\
+& \frac{1}{2}\ A_2 \ \sin^2\theta\ \cos 2 \phi +  A_3 \ \sin \theta\ \cos \phi +     A_4 \ \cos\theta  \nonumber \\
+&     A_5 \ \sin^2\theta \ \sin 2 \phi +     A_6 \ \sin 2 \theta\ \sin \phi +     A_7 \ \sin \theta \ \sin \phi \bigg\}\nonumber.
\end{align}
The dependence of the differential cross section on~$\cos\theta$ and~$\phi$ is analytical and is fully contained in the harmonic polynomials. On the other hand, the dependence on \pTZ, \absyZ\ and~$m$ is entirely contained in the unpolarised cross section, denoted by $\sigma^{U+L}$, and in the \Ai~angular coefficients. Therefore, all hadronic dynamics from the production mechanism are factorised from the decay kinematics in the \Zboson-boson rest frame. This allows the measurement precision to be essentially insensitive to all uncertainties in QCD, quantum electrodynamics (QED), and electroweak (EW) effects related to \Zboson-boson production and decay. In particular, EW~corrections that couple the initial-state quarks to the final-state leptons have a negligible impact (below~0.05\%) at the \Zboson-boson pole. This has been shown for the LEP~precision measurements~\cite{EWLEP1,EWLEP2}, when calculating the interference between initial-state and final-state QED~radiation and more recently for \Zboson-boson measurements at the~LHC~\cite{KKMChh19}. The small fraction of  $\gamma^*$~production in the $Z$-boson pole region and its interference with the $Z$~boson can also be described by~Eq.~(\ref{eq:master2}), but with different coefficients, so the $A_i$~coefficients discussed in this paper are effective coefficients, containing the small $\gamma^*$~contribution at the $Z$~pole.
 
The decomposition is based on a simple and model-independent ansatz: the spin-one nature of the intermediate gauge boson, and the spin one-half nature of the decay leptons, and on the assumption of angular momentum conservation and quantisation. It removes the need for predictions to model accurately the polarisation and decay of the $Z$-boson; only its production properties are of interest for comparison to the measurements. Therefore, any phenomenological interpretation of this measurement avoids the theoretical uncertainties and ambiguities related to spin correlations and to the resummation of fiducial power corrections~\cite{Ebert:2020dfc,Glazov:2020gza,Salam:2021tbm,Alekhin:2021xcu}.
 
One striking example of the advantage of exploiting measurements in the full phase space of the decay leptons is the rapidity dependence of the $Z$-boson transverse momentum spectrum: whereas the fiducial measurements of~Ref.~\cite{ATLAS8TeV} are insensitive to this production property because the fiducial lepton selections essentially remove the rapidity dependence of the spectrum, the measurements reported here probe this dependence very precisely.
The differential angular distributions also provide extra constraints on certain experimental systematic uncertainties since they are not used in the calibration procedures of the detectors. Finally, they have even less sensitivity to theoretical systematic uncertainties than the unfolded~\pTZ\ fiducial measurements.
 
These measurements pave the way for unambiguous interpretations in terms of~QCD. A precise determination of the $Z$-boson transverse momentum spectrum leads to excellent sensitivity to the strong coupling constant at the ~\mZ~scale, while the even more precise rapidity-dependent cross-sections, obtained after integrating over~\pTZ, can be compared directly to state-of-the-art fixed-order predictions with excellent sensitivity to the parton distribution functions. 
 
This paper is structured as follows. Section~\ref{sec:methodology} first presents an overview of the measurement methodology and of the likelihood fit performed to extract the observables of interest from the experimental distributions. Section~\ref{sec:analysis}  describes briefly the data analysis, and Section~\ref{sec:sys} discusses the systematic uncertainties in the measured differential cross-sections. Section~\ref{sec:results} compares the \dsdpTdy~measurements to theoretical predictions combining perturbative resummation with fixed-order calculations, and then compares the~\dsdy\ and the total cross-section measurements to the predictions from QCD N$^3$LO~calculations using different PDF~sets. Finally, Section~\ref{sec:summary} summarises and concludes the paper.
 
\section{Measurement methodology}
\label{sec:methodology}
 
The angular coefficients are extracted from the data by fitting templates of the $\poli$ polynomial terms, defined in~Eq.~(\ref{eq:master2}), to the reconstructed angular distributions using $8 \times 8$~bins in ($\cos\theta,\phi$)~space (see~Ref.~\cite{ATLASAi} for a detailed description). Each template is normalised by a free parameter for its corresponding polynomial coefficient~$\Ai$ and a common parameter representing the unpolarised cross section. The polynomial $P_8 = 1+\cos^2\theta$ in Eq.~(\ref{eq:master2}) is only normalised by the parameter for the unpolarised cross-section. All the angular coefficients together with the corresponding unpolarised cross section parameters are measured in each of the analysis bins in (\pTZ,~\absyZ)~space.
 
In the absence of selections for the final state leptons, the angular distributions in the gauge boson rest frame are defined by its polarisation. In the presence of selection criteria applied to the leptons, the distributions are sculpted by kinematic effects, and can no longer be described by the sum of the nine $\poli$~polynomials as in Eq.~(\ref{eq:master2}). Templates of the $\poli$~terms are therefore constructed to account for this, which requires fully simulated signal Monte-Carlo~(MC) samples to model the acceptance, efficiency, and migration of events. Reference~$\Ai$ coefficients are extracted in a simple way from the predicted shapes of the angular distributions in the full phase space of the decay leptons, using the orthogonality of the $P_i$~polynomials (see~Ref.~\cite{arXiv9406381} describing the underlying moment method used to extract the coefficients). Together with the reference unpolarised cross-section, they are used in a folding procedure based on the signal MC~simulation. The folded polynomial templates (or simply templates) are built in ($\cos\theta,\phi$)~space for each of the nine original polynomials and for each of the measurement bins in (\pTll,~\yll)~space. They are then used to extract the angular coefficients and the unpolarised cross section in the full phase space of the leptons from $Z$-boson~decay. The observables~\mll,~\pTll, and~\yll, which are defined using reconstructed lepton pairs, as described in Section~\ref{sec:analoverview}, are to be distinguished from~$m$,~\pTZ~and~\absyZ, which are defined at generator level using lepton pairs at the Born~level.
 
A likelihood is built from the nominal templates and the varied templates reflecting the systematic uncertainties, which are represented by two categories of nuisance parameters~(NP), $\beta$~and~$\gamma$. The first category,~$\beta$, represents experimental and theoretical uncertainties. Each $\beta^m$ in the set $\beta~=~\left\{\beta^1,...,\beta^M\right\}$ is constrained by a unit Gaussian probability density function,~$G(0|\beta^m, 1)$, and linearly interpolates between the nominal and varied templates. The second category,~$\gamma^n$, represents systematic uncertainties from the limited size of the MC~signal and background samples, which are constrained by Poisson probability density functions,~$P(N_{\mathrm{eff}}^{\mathrm{n}}|\gamma^n N_{\mathrm{eff}}^{\mathrm{n}})$ where $N_{\mathrm{eff}}^{\mathrm{n}}$~is the effective number of MC~events in bin~$n$, in each of the $N_{\mathrm{bins}}$~=~22528~bins of the measurement. After including these auxiliary parameters, and after all signal and background templates (see Section~\ref{sec:analysis} for details of the samples) are summed over (with their respective normalisations), the expected number of events $N_{\mathrm{exp}}^n$ in bin~$n$ of the measurement can be written as:
\begin{equation}
\label{eq:Nevts}
N_{\mathrm{exp}}^{n}(A,\sigma^{U+L},\beta,\gamma) = \left\{\displaystyle\sum_{j}\sigma^{U+L}_{j}\times L\times\left[t_{8j}^{n}(\beta) + \displaystyle\sum_{i=0}^{7}A_{ij}\times t_{ij}^{n}(\beta)\right ] \right\}\times\gamma^n + \displaystyle\sum_{B}^{\mathrm{bkgs}}T_{B}^{n}(\beta),
\end{equation}
where:
\begin{itemize}
\item index $i$ runs over the eight angular coefficients and the corresponding $\poli$~polynomials, while index~$j$ runs over all 352~bins in (\pTZ,\absyZ) space
\item $A$ is the set of all angular coefficients, $A_{ij}$
\item $\sigma^{U+L}$ is the set of all unpolarised cross sections, $\sigma_{j}^{U+L}$
\item $\beta$ is the set of all Gaussian-constrained nuisance parameters representing the systematic uncertainties
\item $\gamma$ is the set of all Poisson-constrained nuisance parameters representing the statistical uncertainties in the simulated samples and in the background estimates
\item $t$ is the set of all signal $P_i$ polynomial templates, $t_{ij}$
\item $T_{B}$ is the set of background templates, where the sum runs over all background sources
\item $L$ is the total integrated luminosity.
\end{itemize}
The summation over index $j$ accounts for the contributions from all analysis bins at generator level that migrate into other analysis bins at the reconstruction level. The likelihood is then constructed as a product of Poisson probabilities across all $N_{\mathrm{bins}}$ and of auxiliary constraints for each nuisance parameter $\beta^m$:
\begin{equation}
\label{eq:likelihood}
\mathcal{L}(A,\sigma^{U+L},\theta|N_{\mathrm{obs}}) = \displaystyle\prod_{n}^{N_{\mathrm{bins}}}\left\{P(N_{\mathrm{obs}}^{n}|N_{\mathrm{exp}}^{n}(A,\sigma^{U+L},\theta))P(N_{\mathrm{eff}}^{n}|\gamma^n N_{\mathrm{eff}}^{\mathrm{n}})\right\}\times\displaystyle\prod_{m}^{M}G(0|\beta^m,1).
\end{equation}
A profile likelihood ratio method is used to extract the best fit values of the parameters of interest~(POIs) and their uncertainties. The POIs include the angular coefficients~\Aij\ and the cross sections~\sigmaj$^{U+L}$\ in each measurement bin. This procedure extends that presented in~Ref.~\cite{ATLASAi}, in which only the angular coefficients were extracted as a function of~\pTZ\ and~\absyZ\ without any fit constraints based on the \sigmaj$^{U+L}$~POIs. The results from the full fit focus on the differential cross-section results in each channel and each measurement bin. The results of the fit for the angular coefficients involving only central leptons are compatible with those of~Ref.~\cite{ATLASAi} within the uncertainties resulting from the slightly different selections and calibrations applied here. As explained in~Section~\ref{sec:analoverview}, the analysis of the forward electrons is significantly improved here with respect to that documented in~Ref.~\cite{PERF-2016-01}. As a result, the measurements of the \eeCF~angular coefficients reported in this paper are published in~HEPDATA and supersede those of~Ref.~\cite{ATLASAi}.

\section{Analysis}
\label{sec:analysis}

\newcommand{\AtlasCoordFootnote}{
ATLAS uses a right-handed coordinate system with its origin at the nominal interaction point (IP) in the centre of the detector and the $z$-axis along the beam pipe. The $x$-axis points from the IP to the centre of the LHC ring, and the $y$-axis points upwards. Cylindrical coordinates $(r,\phi)$ are used in the transverse plane, $\phi$ being the azimuthal angle around the $z$-axis. The pseudorapidity is defined in terms of the polar angle $\theta$ as $\eta = -\ln \tan(\theta/2)$. Angular distance is measured in units of $\Delta R \equiv \sqrt{(\Delta\eta)^{2} + (\Delta\phi)^{2}}$.}

\subsection{ATLAS detector}
 
The ATLAS experiment~\cite{ATLASexp} at the LHC is a multipurpose particle detector with a forward--backward symmetric cylindrical geometry and a near $4\pi$ coverage in solid angle.\footnote{\AtlasCoordFootnote} It consists of an inner tracking detector, electromagnetic (EM) and hadronic calorimeters, and a muon spectrometer. The inner detector provides precision tracking of charged particles in the pseudorapidity range $|\eta| < 2.5$. This region is matched to a high-granularity EM sampling calorimeter covering the pseudorapidity range $|\eta| < 3.2$ and a coarser granularity calorimeter up to $|\eta| = 4.9$. A hadronic calorimeter system covers the entire pseudorapidity range up to  $|\eta| = 4.9$. The muon spectrometer provides triggering and tracking capabilities in the range $|\eta| < 2.4$ and $|\eta| < 2.7$, respectively.
 
A first-level trigger is implemented in hardware, followed by two software-based trigger levels that together reduce the accepted event rate to \SI{400}{\hertz} on average. An extensive software suite~\cite{ATL-SOFT-PUB-2021-001} is used in the reconstruction and analysis of real and simulated data, in detector operations, and in the trigger and data acquisition systems of the experiment.
 
\begin{table}
\small
\setlength\extrarowheight{3pt}
\begin{center}
\begin{tabular}{lcc}
\toprule
Process                        & Generator   &PDF  \\ 
\midrule
$Z/\gamma^* \rightarrow \ell \ell$ & \POWHEGBOX \cite{Pow2,Frixione:2007vw,Alioli:2010xd} +~\PYTHIA8 \cite{arXiv:0710.3820}    &  {\textsc CT10 NLO} \cite{Lai:2010vv} \\ 
\midrule
$Z/\gamma^* \rightarrow \tau \tau$ & \SHERPA  \cite{Gleisberg:2008ta,Hoeche:2009rj,Gleisberg:2008fv,Schumann:2007mg}  & {\textsc CT10 NLO}   \\
$t\bar{t}$                    & \POWHEGBOX +~\PYTHIA6 \cite{Sjostrand:2006za}   & {\textsc CT10 NLO}  \\
Single top quark ($Wt$ channel)         & \POWHEGBOX +~\PYTHIA8   & {\textsc CT10 NLO}     \\
Dibosons                           & \HERWIG   \cite{Corcella:2000bw}          & {\textsc CTEQ6L1}             \\
$\gamma\gamma\to \ell\ell$        & \PYTHIA8            & {\textsc MRST2004QED NLO}  \cite{Martin:2004dh}   \\
\bottomrule 
\end{tabular}
\end{center}
\caption{ MC samples used to estimate the signal and backgrounds in the analysis. }
\label{Tab:Zall_MC_samples}
\end{table}

\subsection{Data and Monte Carlo samples}
\label{sec:dataMC}
 
The data were collected by the ATLAS detector in 2012 at a centre-of-mass energy of $\sqrt{s} = 8$~TeV, and correspond to an integrated luminosity of~20.2~fb$^{-1}$. The mean number of additional $pp$ interactions per bunch crossing (pile-up events) in the data set is approximately~20.
 
The event generators used to produce the $Z/\gamma^*\rightarrow \ell\ell$ signal events and the backgrounds estimated from simulation are listed in Table~\ref{Tab:Zall_MC_samples}. The baseline \POWHEGBOX~(v1/r2129) signal sample~\cite{Pow2,Frixione:2007vw,Alioli:2010xd}, using the {\textsc CT10 NLO} set of~PDFs~\cite{Lai:2010vv}, is interfaced to \PYTHIA~8~(v.8.170)~\cite{arXiv:0710.3820} with the AU2~set of tuned parameters~\cite{ATL-PHYS-PUB-2012-003} to simulate the parton shower, hadronisation and underlying event, and to Photos~(v2.154)~\cite{Golonka:2005pn} to simulate QED~final-state radiation~(FSR) in the $\Zboson$-boson decay. The simulated line-shape of the $Z$-boson is reweighted to account for the running width and for mass-dependent NNLO~QCD~\cite{FEWZ}~effects, following the recipe described in~Ref.~\cite{STDM-2016-04}. This is a percent-level effect within the mass range used by the measurement, and provides the best possible leading-order EW~prediction of the line-shape of the Z boson. The number of events available in the baseline \POWHEGBOX+~\PYTHIA8 signal sample corresponds to approximately four~times that in the data.

Backgrounds from EW (diboson, $Z/\gamma^* \rightarrow \tau \tau$, and $\gamma\gamma\to \ell\ell$ production) and top-quark (production of top-quark pairs and of single top quarks) processes are evaluated from the MC samples listed in Table~\ref{Tab:Zall_MC_samples}. All MC~samples are processed through a full ATLAS detector simulation~\cite{SOFT-2010-01}, based on \textsc{Geant4}~\cite{Agostinelli:2002hh}, and reconstructed with the same software as that used for the data.
Pile-up events, occurring in the same and neighbouring bunch crossings are simulated and overlaid at the detector hit level on top of the hard-scattering process from the MC~simulation.

\subsection{Analysis overview and event selection}
\label{sec:analoverview}
 
For this paper, a central lepton (electron or muon) is one found in the region $|\eta| < 2.4$ (excluding, for electrons, the electromagnetic calorimeter barrel/end-cap transition region~$1.37 < |\eta| < 1.52$), while a forward electron is one found in the region $2.5 < |\eta| < 4.9$ (excluding a transition region~$3.00 < |\eta| < 3.35$ between the end-cap and forward calorimeters). The analysis is split into three orthogonal channels that are analysed independently, and then combined for the last stages of the analysis, after verifying their compatibility.
 
The \eeCC~channel consists of candidate events with two central electrons, obtained using a logical~OR of a dielectron trigger, requiring two electron candidates, each having $\pT~>~12$~GeV, with two high-\pT\ single-electron triggers, the main one corresponding to a \pT~threshold of 24~GeV. Electron candidates are required offline to have $\pT~>~20$~GeV and~$\abseta < 2.4$, and are reconstructed from clusters of energy in the electromagnetic calorimeter matched to inner detector tracks. The electron candidates must satisfy a set of ``medium'' selection criteria~\cite{PERF-2016-01}, which have been optimised for the level of pile-up present in the 2012 data.
 
The \mmCC~channel consists of candidate events with two central muons, obtained using a logical~OR of a dimuon trigger requiring two muon candidates with~$\pT > 18$~GeV and 8 GeV, respectively, and of two high-$\pT$ single-muon triggers, the main one corresponding to a \pT~threshold of 24~GeV. Muon candidates are required offline to have $\pT > 20$~GeV and~$\abseta < 2.4$, and are identified as tracks in the inner detector, which are matched and combined with track segments in the muon spectrometer~\cite{PERF-2014-05hack}. Track-quality, and longitudinal and transverse impact-parameter requirements are imposed to suppress backgrounds, and to ensure that the muon candidates originate from a common primary $pp$~interaction vertex.
 
The \eeCF~channel consists of candidate events with one central and one forward electron, obtained using the logical~OR of the two high-\pT\ single-electron triggers used for the \eeCC~events, as described above. The central electron candidate is required to have $\pT > 25$~GeV and~$\abseta < 2.4$. Because the expected background from multijet events is larger in this channel than in the \eeCC~channel, the central electron candidate is required in addition to satisfy a set of ``tight'' selection criteria~\cite{PERF-2016-01}, including an explicit isolation requirement, which are optimised for the level of pile-up observed in the 2012 data. The forward electron candidate is required to have $\pT > 20$~GeV and to satisfy a set of ``medium'' selection criteria, based only on the shower shapes in the calorimeters, since this region is outside the acceptance of the inner detector~\cite{PERF-2013-03}.
 
In the case of the \eeCF~channel, the analysis of the forward electron candidates, in terms of the data quality, alignment, identification criteria, and calibration of the electromagnetic compartments of the calorimeters outside the acceptance of the inner detector, was improved significantly from those documented in~Ref.~\cite{PERF-2016-01}. A few examples illustrating these improvements are:
\begin{itemize}
\item misalignments of a few~mm between the inner wheels of the electromagnetic end-cap calorimeters and the inner detector and a rotation of a few~mrad between one of the forward calorimeters and the inner detector were found and corrected for when studying the azimuthal angle difference between the central and forward electrons. As a result, the two angular coefficients~($A_1,A_6$), for which the fully two-dimensional ($\cos\theta,\phi$) measurement is necessary, can now be measured for the \eeCF~channel, and are found to be in agreement with the SM~predictions.
\item an azimuthal inter-calibration of the forward electrons was performed to improve the baseline calibration. This corrected for large inhomogeneities of up to~10--15\% in the energy response due to material from inner-detector services in front of the calorimeters not accounted for correctly in the simulation, and for high-voltage problems in some cells, which were inadequately corrected for online during data-taking.
\item the largest improvement was obtained by adjusting the simulated lateral shower shapes in the calorimeters to data and then calibrating carefully each region of the calorimeters. This resulted in both improved efficiencies for the forward electrons in certain regions of phase space and improved energy calibration.
\end{itemize}
As a result, the p-value for the agreement between the combined \eeCC~and~\mmCC~channels and the \eeCF~channel in the \absyZ~range where they overlap, $1.6 < \absyZ < 2.4$, was improved from~$ < 10^{-4}$ to~3\% (see~Fig.~\ref{compatibility} in~Section~\ref{sec:compatibility}).
 
In each channel, the events are required to contain exactly two lepton candidates satisfying the criteria described above. For the central-central channels, where the charges of both leptons are measured, the two lepton candidates are required to be of opposite charge.
 
\begin{figure}[h]
\begin{center}
\includegraphics[width=0.49\textwidth]{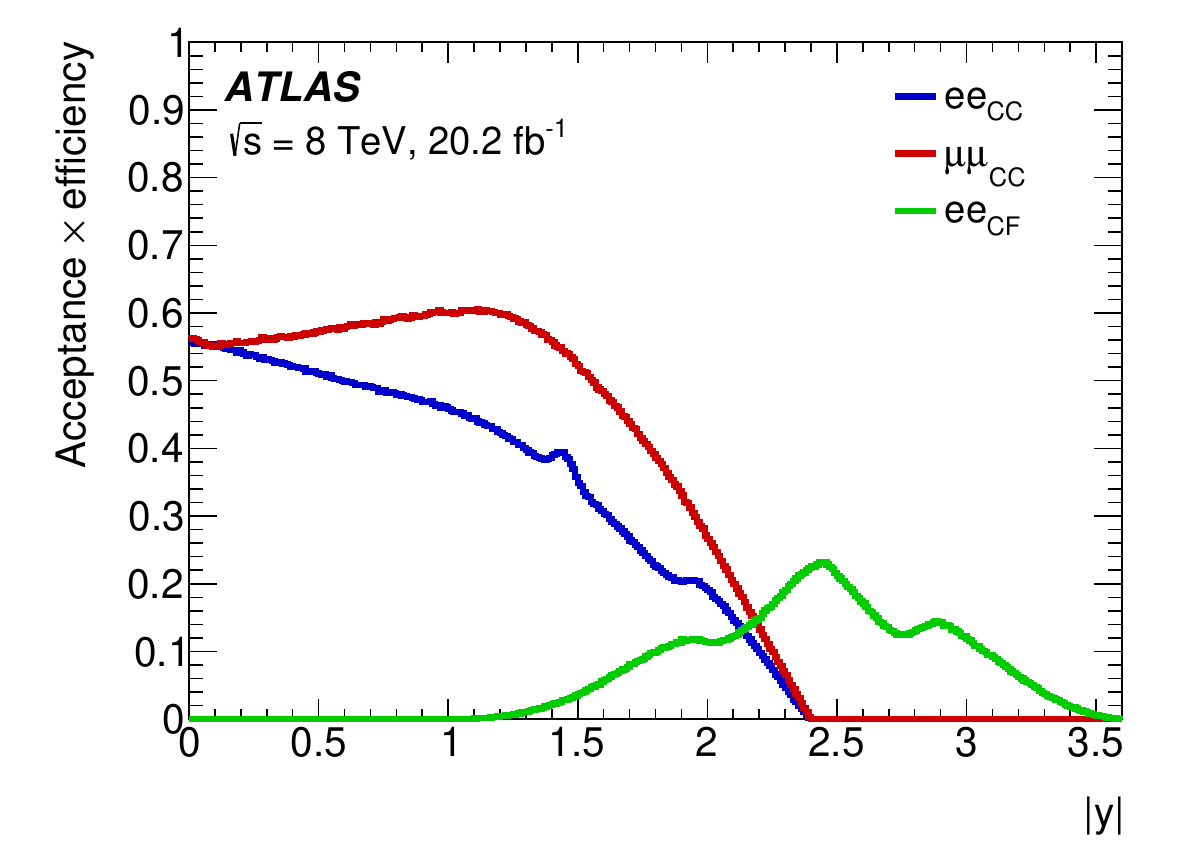}
\includegraphics[width=0.49\textwidth]{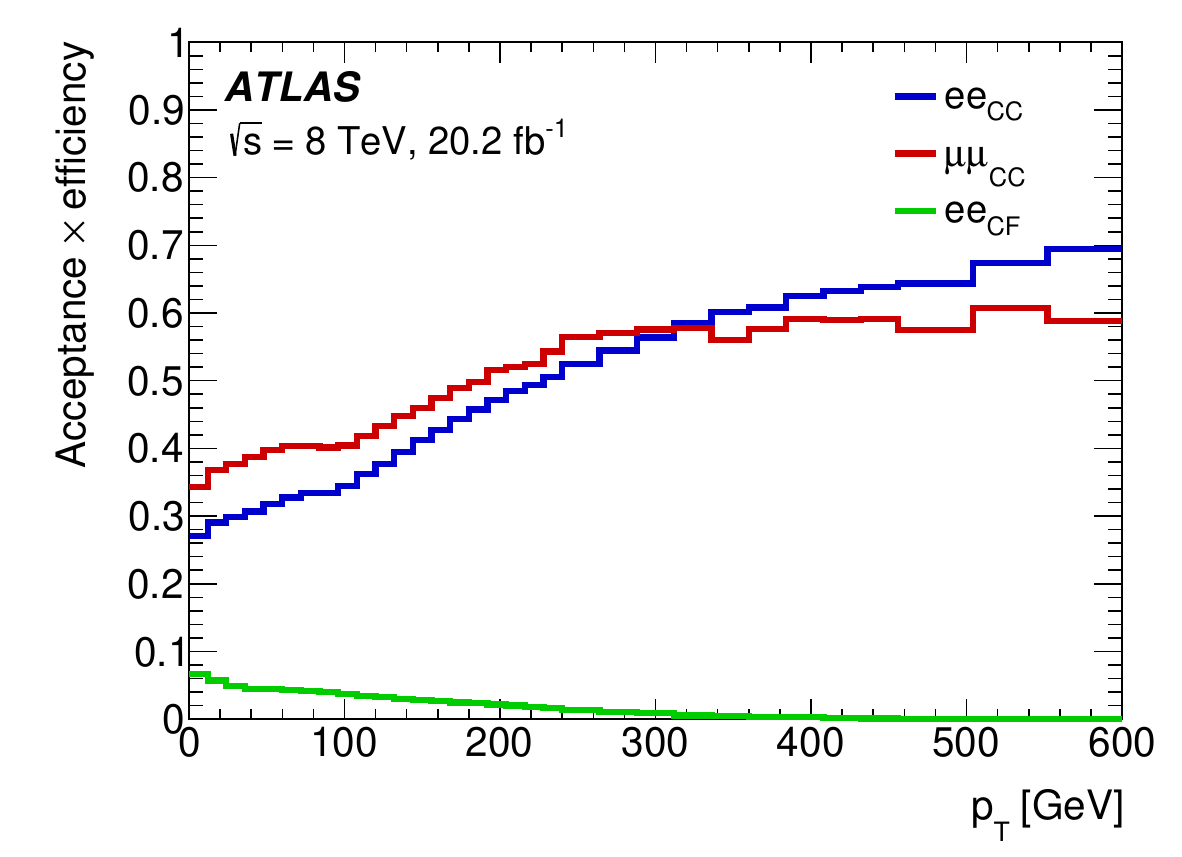}\\
\end{center}
\vspace*{-0.5cm}
\caption{Expected $Z$-boson acceptance times selection efficiencies as a function of~\absyZ~(left) and~\pTZ~(right) for the \eeCC, \mmCC\ and \eeCF\ analysis channels. }
\label{fig:expEvtAccEff}
\end{figure}
 
As described in Section~\ref{sec:Intro}, this analysis is focused on the $Z$-boson pole region, and the lepton pair is required to have an invariant mass,~\mll, within a window around the $Z$-boson mass, $80 < \mll < 100$~GeV. The simulated events are required to satisfy the same selection criteria as the data, after applying small corrections to account for the differences between data and simulation in terms of reconstruction, identification and trigger efficiencies and of energy scale and resolution for electrons and muons~\cite{PERF-2013-03,PERF-2013-05hack,PERF-2014-05hack}. All simulated events are reweighted to match the distributions observed in data for the level of pile-up and for the primary vertex longitudinal position.

The measurements are binned in \pTZ\ and \absyZ, with fine bins in~\pTZ\ and coarse bins in~\absyZ, as follows:
\begin{enumerate}
\item \eeCC\ and \mmCC\ channels:
\begin{itemize}
\item 23 bins in \pTZ\ with bin boundaries $\{$0, 2.5, 5.0, 8.0, 11.4, 14.9, 18.5, 22.0, 25.5, 29.0, 32.6, 36.4, 40.4, 44.9, 50.2, 56.4, 63.9, 73.4, 85.4, 105.0, 132.0, 173.0, 253.0, 4000$\}$ \GeV,
\item 6 bins in \absyZ\ with bin boundaries $\{$0, 0.4, 0.8, 1.2, 1.6, 2.0, 2.4$\}$;
\end{itemize}
\item \eeCF\ channel:
\begin{itemize}
\item 19 bins in \pTZ\ with bin boundaries $\{$0, 2.5, 5.0, 8.0, 11.4, 14.9, 18.5, 22.0, 25.5, 29.0, 32.6, 36.4, 40.4, 44.9, 50.2, 56.4, 63.9, 73.4, 85.4, 105.0$\}$ \GeV,
\item 4 bins in \absyZ\ with bin boundaries $\{$1.6, 2.0, 2.4, 2.8, 3.6$\}$.
\end{itemize}
\end{enumerate}
The choice of bin boundaries in~\pTZ\ is the result of an optimisation with respect to the limited resolution of the measurements at low~\pTZ\ and the limited statistics at high~\pTZ.
 
The angular coefficients and unpolarised cross-section results are extracted taking into account the correlations and migrations between the measurement bins. Figure~\ref{fig:expEvtAccEff} shows the expected products of the acceptance and the selection efficiency, defined as the ratio of the number of selected events to the number of events in the full decay lepton phase space, as functions of both~\absyZ\ and~\pTZ. The shape of this product of acceptance and efficiency in~\absyZ\ results in part from the differences in reconstruction and identification efficiencies between central electrons and muons. The \eeCF\ channel covers a higher range in~\absyZ\ and overlaps with the central-central channels for~$1.6 < \absyZ < 2.4$. In each of the analysis measurement bins, two-dimensional ($\cos \theta,\phi$) angular distributions are computed (eight equal-sized bins in each observable) and they serve as the basis for the simultaneous extraction of all the angular coefficients and of the unpolarised cross-sections, following the methodology described in~Section~\ref{sec:methodology}.

\subsection{Background estimates and signal yields}
 
Table~\ref{tab:yields} shows the event yields and breakdowns of the fraction of events originating from background sources, for the \eeCC,~\mmCC, and~\eeCF\ channels, respectively. The numbers are shown in each \absyZ~analysis bin integrated over~\pTZ. The backgrounds are divided into three major groups:
\begin{itemize}
\item the backgrounds containing a prompt isolated lepton, from top~($t\bar t$ and~$Wt$), diboson, $Z\to\tau\tau$, and $\gamma\gamma \to ll$ processes, estimated from simulation samples (see Section~\ref{sec:dataMC}),
\item the multijet and $W$+jet backgrounds, estimated from data using a method very similar to that described in~Ref.~\cite{ATLASAi}, and
\item the non-fiducial \Zboson\ background, which is estimated from simulation and consists almost entirely of events outside the full mass range considered for the analysis at generator level, but passing the selection cuts at reconstruction level owing to migrations.
\end{itemize}
The background contamination from other processes for the~\eeCC\ and \mmCC~channels is very small and amounts to about~0.3\%. The multijet background in the \eeCF~channel is larger, amounting to about~1.0\%. The non-fiducial \Zboson\ background amounts to about~1\% in all channels. Templates of the angular distributions of the sum of all these backgrounds are used in the fit to extract the angular coefficients, as described in~Section~\ref{sec:methodology}.
 
\subsection{Angular distributions}
 
\begin{figure}[htb!]
\begin{center}
\includegraphics[width=0.49\textwidth]{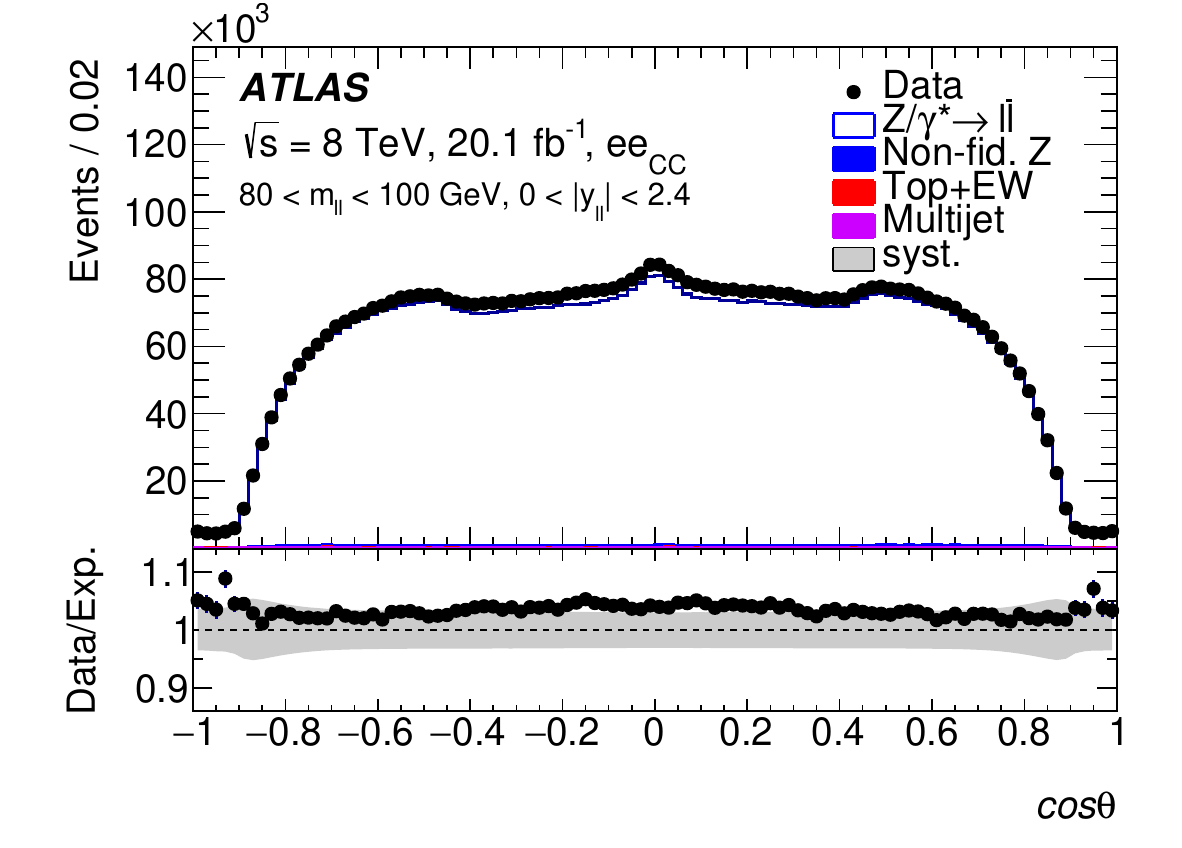}
\includegraphics[width=0.49\textwidth]{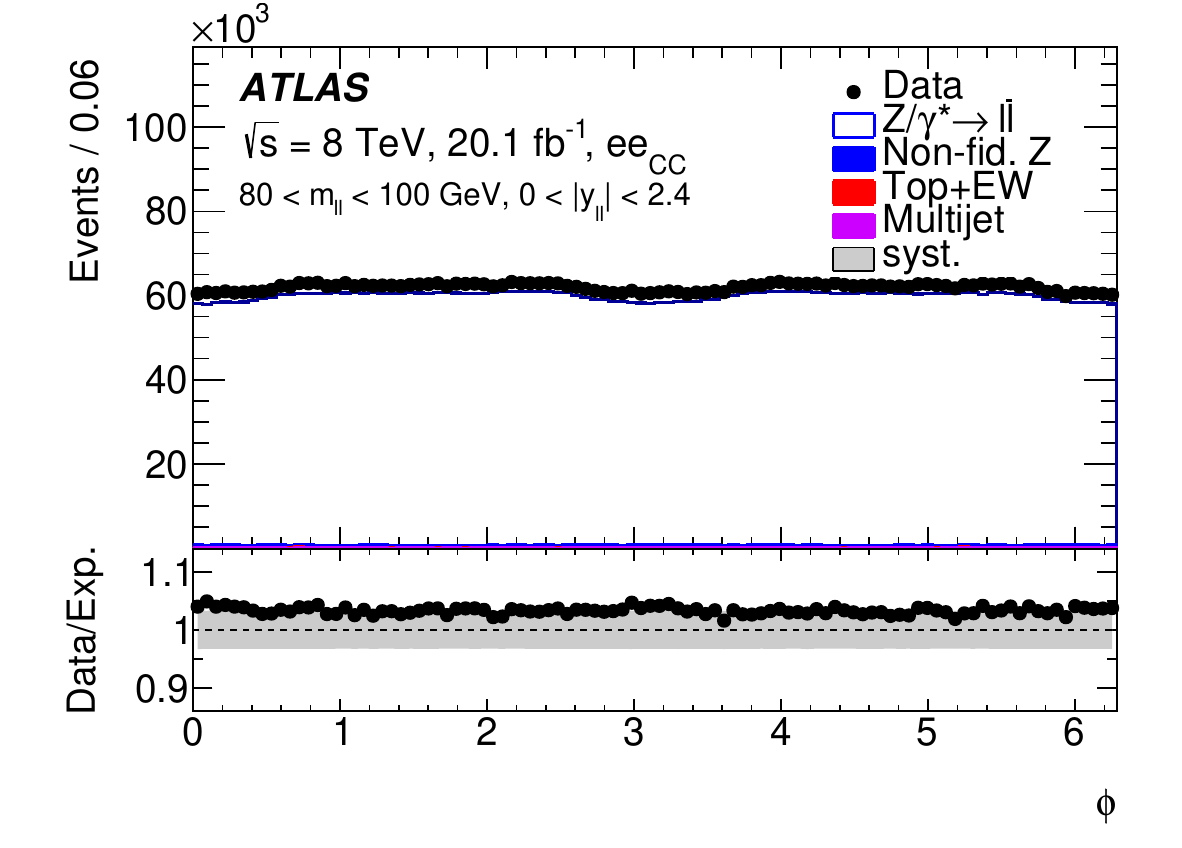}
\includegraphics[width=0.49\textwidth]{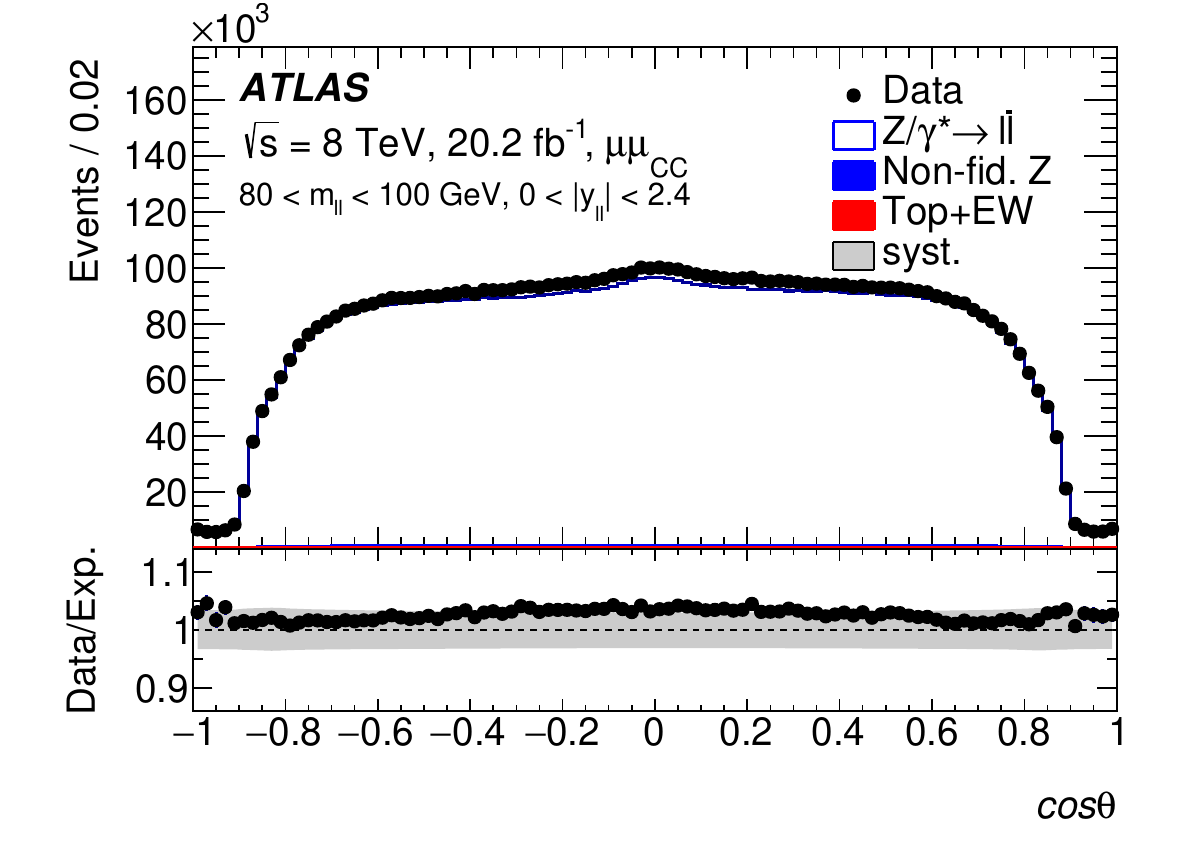}
\includegraphics[width=0.49\textwidth]{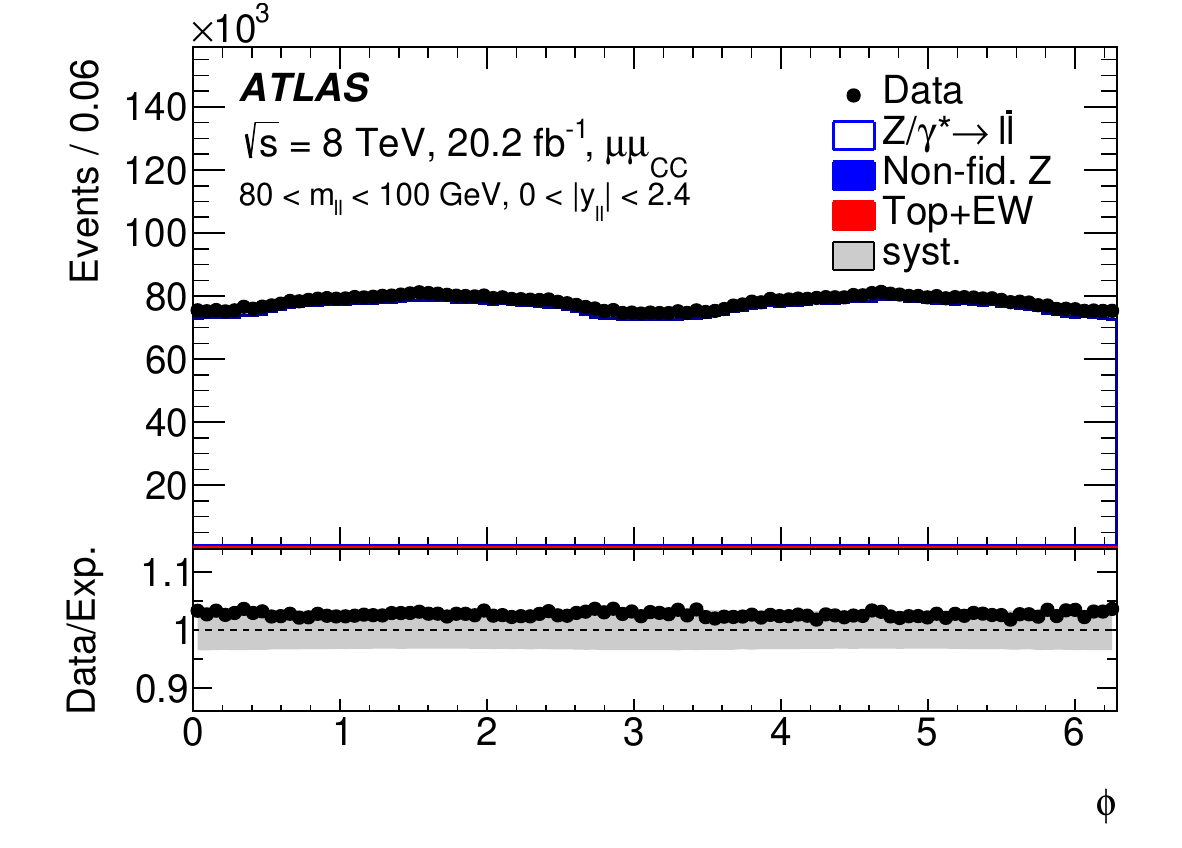}
\includegraphics[width=0.49\textwidth]{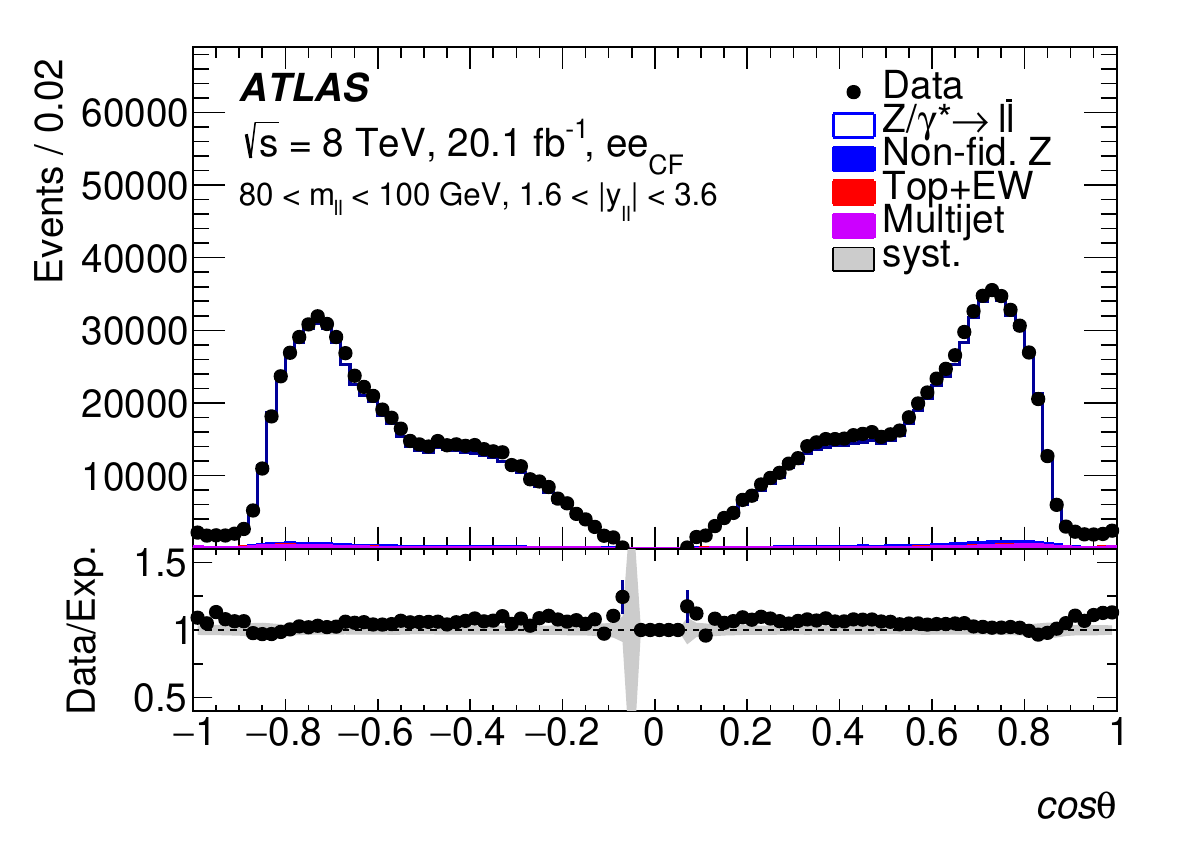}
\includegraphics[width=0.49\textwidth]{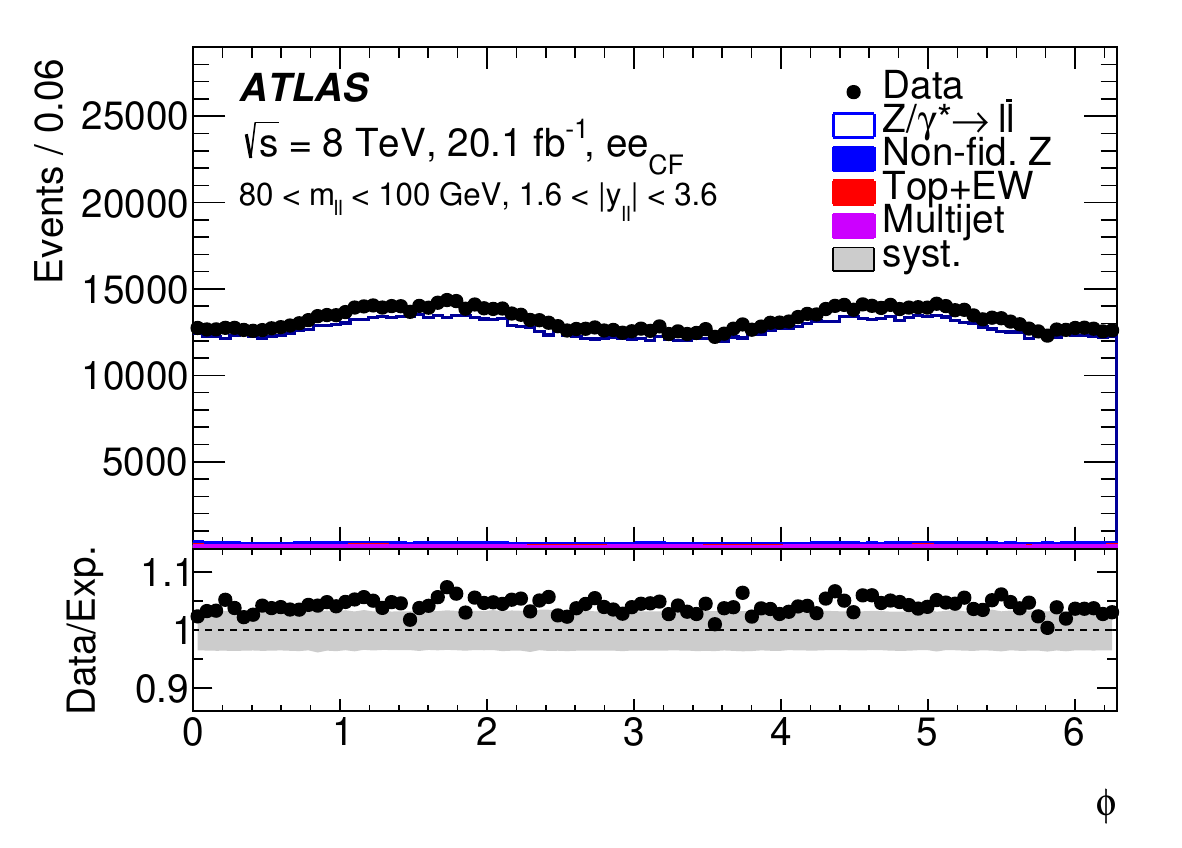}
\caption{
The cos$\theta$ (left) and $\phi$ (right) angular distributions, integrated over~\pTll and~\yll, for the \eeCC~(top), \mmCC~(middle) and \eeCF~(bottom) channels. The pre-fit expected distributions are shown separately for the signal and the different small background sources contributing to each channel. The grey band represents the experimental systematic uncertainties in the total expected yields from signal plus background. The small normalisation differences between the data and the MC~distributions observed at the level of a few percent are compatible with the combination of uncertainties in integrated luminosity and signal cross section.
\label{costheta/phi}}
\end{center}
\end{figure}

The criteria described above are applied to the data, leading to totals of 6.2~(\eeCC), 7.8~(\mmCC), and 1.3~(\eeCF) million selected events (see Table~\ref{tab:yields} for a more detailed breakdown as a function of~\yll). The reconstructed differential angular distributions, integrated over~\pTll\ and~\yll, are shown in~Fig.~\ref{costheta/phi} for the three channels. Small normalisation differences between the data and the MC~distributions are observed at the level of a few percent, compatible with the combination of uncertainties in integrated luminosity and signal cross section. The measurement of the angular coefficients is, however, independent of the normalisation between data and simulation in each measurement bin. Overall, the agreement in shape between the observed and predicted distributions of~$\cos\theta$ and~$\phi$ is good.

\begin{table}[h]
\setlength\extrarowheight{2pt}
\begin{center}
\caption{Data yields integrated over~\pTll\ in each \absyll~analysis bin along with estimated background fractions in~\% for the \eeCC, \mmCC, and \eeCF~ channels.
\label{tab:yields}}
\vspace{5mm}
\begin{tabular}{ccccc}
\toprule 
& \multicolumn{4}{c}{\eeCC\ channel} \\
\midrule 
\absyll\ range  & Data & Top+EW & Multijets & Non-fiducial $Z$ \\
& (yield) & (\%) & (\%) &  (\%) \\
\midrule 
0 -- 0.4 & 1 573 411 & 0.2 & 0.1 &  0.8 \\
0.4 -- 0.8 & 1 441 923 & 0.2 & 0.1 & 0.9 \\
0.8 -- 1.2 & 1 285 026 & 0.2 & 0.2 &  0.9 \\
1.2 -- 1.6 & 1 025 043 & 0.2 & 0.1 &  0.9 \\
1.6 -- 2.0 & 621 017 & 0.1 & 0.2 &   1.0 \\
2.0 -- 2.4 & 248 261 & 0.1 & 0.3 &  0.9 \\
\midrule 
$\absyll~< 2.4$ & 6 194 681 & 0.2 & 0.1 & 0.9 \\
\bottomrule 
& \multicolumn{4}{c}{\mmCC\ channel} \\
\midrule 
\absyll\ range& Data & Top+EW & Multijets & Non-fiducial $Z$ \\
& (yield) & (\%) & (\%) &  (\%) \\
\midrule 
0 -- 0.4 & 1 617 521 & 0.3 & $<$~0.1 & 0.7 \\
0.4 -- 0.8 & 1 657 881 & 0.2 & $<$~0.1 & 0.7 \\
0.8 -- 1.2 & 1 680 981 & 0.2 & $<$~0.1 & 0.8 \\
1.2 -- 1.6 & 1 503 977 & 0.2 & $<$~0.1 & 0.8 \\
1.6 -- 2.0 & 1 010 432 & 0.2 & $<$~0.1 & 0.9 \\
2.0 -- 2.4 & 335 788 & 0.2 & $<$~0.1 & 0.9 \\
\midrule 
$\absyll~< 2.4$ & 7 806 580 & 0.2 & $<$~0.1 & 0.8 \\
\bottomrule 
& \multicolumn{4}{c}{\eeCF\ channel} \\
\midrule 
\absyll\ range & Data & Top+EW & Multijets & Non-fiducial $Z$ \\
& (yield) & (\%) & (\%) &  (\%) \\
\midrule 
1.6 -- 2.0 & 245 869 & 0.2 & 1.3 & 1.0 \\
2.0 -- 2.4 & 388 183 & 0.1 & 0.9 & 1.1 \\
2.4 -- 2.8 & 391 405 & 0.1 & 0.7 & 1.1 \\
2.8 -- 3.6 & 228 867 & $<$~0.1 & 1.1 & 1.2 \\
\midrule 
$1.6 < \absyll~< 3.6$ & 1 254 324 & 0.1 & 1.0 & 1.1 \\
\bottomrule 
\end{tabular}
\end{center}
\end{table}
 
\clearpage
 
\begin{table}[htb]
\begin{center}
\caption{ \label{SysTotX} Examples of breakdown of relative uncertainties in the measured absolute differential cross-sections, as obtained from the full fit. The values are given in~\%\ for the region near the peak of the \pTZ~distribution and are shown for two \yZ~bins in the case of the~\eeCC\ and \mmCC~channels and for one \yZ~bin in the case of the \eeCF~channel. Also shown are the total uncertainties. The uncertainty of~1.8\% in the integrated luminosity is not included. }
\vspace{2mm}
\resizebox{\columnwidth}{!}{
\begin{tabular}{lccccc}
\toprule 
Uncertainty source & \eeCC\ channel & \mmCC\ channel & \eeCC\ channel & \mmCC\ channel & \eeCF\ channel \\
& (\%) & (\%) &(\%) & (\%) &(\%) \\% \hline
\midrule
Rapidity range & $0 < \absyZ\ < 0.4$ & $0 < \absyZ\ < 0.4$ & $1.2 < \absyZ\ < 1.6$ & $1.2 < \absyZ\ < 1.6$ & $2.4 < \absyZ\ < 2.8 $  \\ 
\midrule
Data stat. & 0.5  & 0.5  & 0.8  & 0.6  & 2.7   \\
MC stat. & 0.3  & 0.3  & 0.4  & 0.3  & 1.5   \\
Leptons & 0.2  & 0.3  & 0.2  & 0.4  & 0.6   \\
Background  & $<$ 0.1  & $<$ 0.1  & $<$ 0.1  & $<$ 0.1  & $<$ 0.1    \\
PDF & $<$ 0.1  & $<$ 0.1 & $<$ 0.1  &  $<$ 0.1 & $<$ 0.1   \\
\midrule
Total & 0.6  & 0.6  & 1.0  & 0.8 & 3.0  \\
\bottomrule 
\end{tabular}
}
\end{center}
\end{table}

\section{Systematic uncertainties}
\label{sec:sys}
 
\begin{figure}[htb!]
\begin{center}
\includegraphics[width=\textwidth]{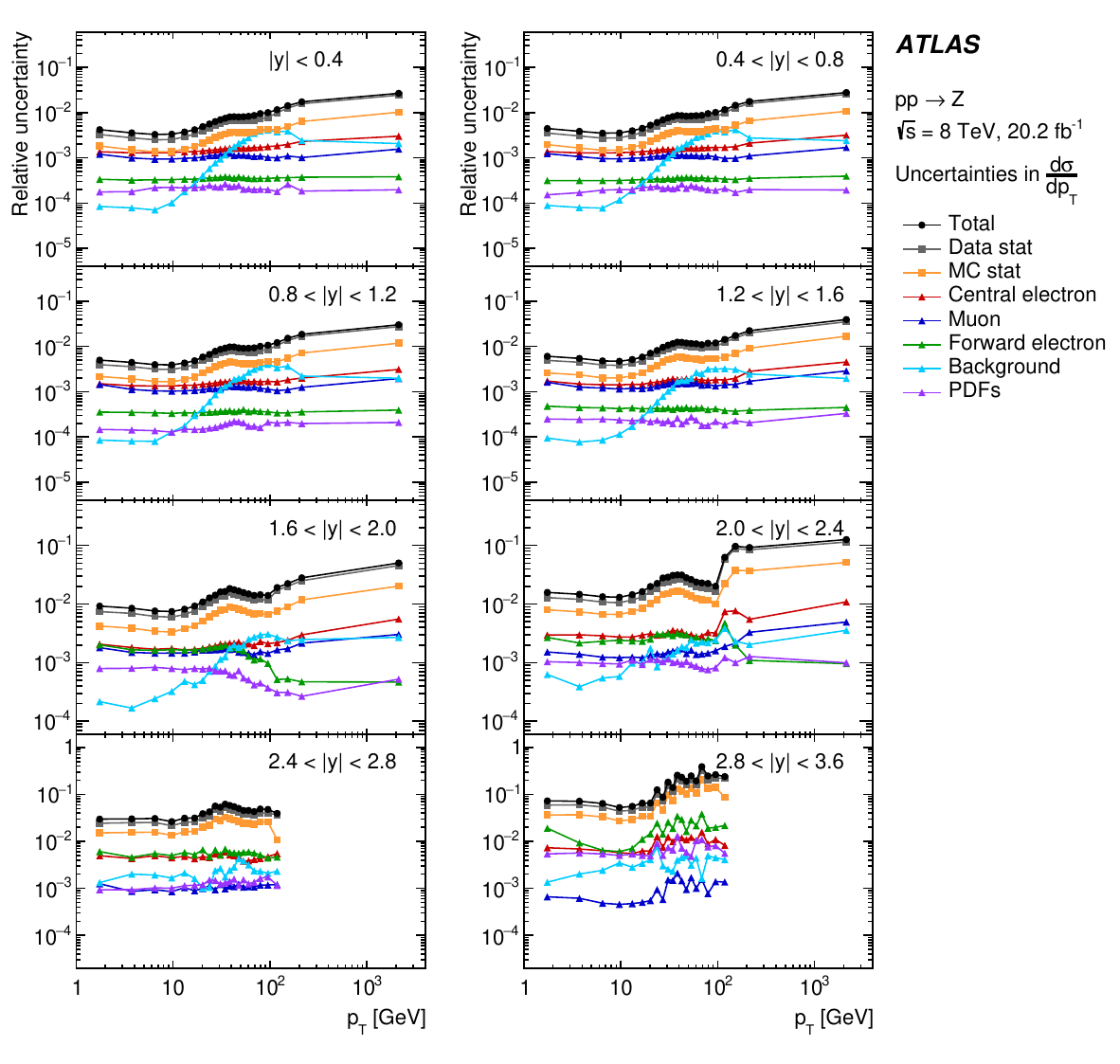}
\caption{
Breakdown of relative uncertainties in the measured absolute differential cross-sections,~\dsdpTdy, as a function of~\pTZ. The values are shown, as obtained from the full fit and for each \absyZ~bin, for the combination of the~\eeCC\ and \mmCC~channels for $\absyZ < 1.6$, for the combination of all three channels for $1.6 < \absyZ < 2.4$, and for the \eeCF~channel alone for $2.4 < \absyZ < 3.6$. The uncertainty of~1.8\% in the integrated luminosity is not included.
\label{uncpT}}
\end{center}
\end{figure}

\begin{figure}[htb!]
\begin{center}
\includegraphics[width=\textwidth]{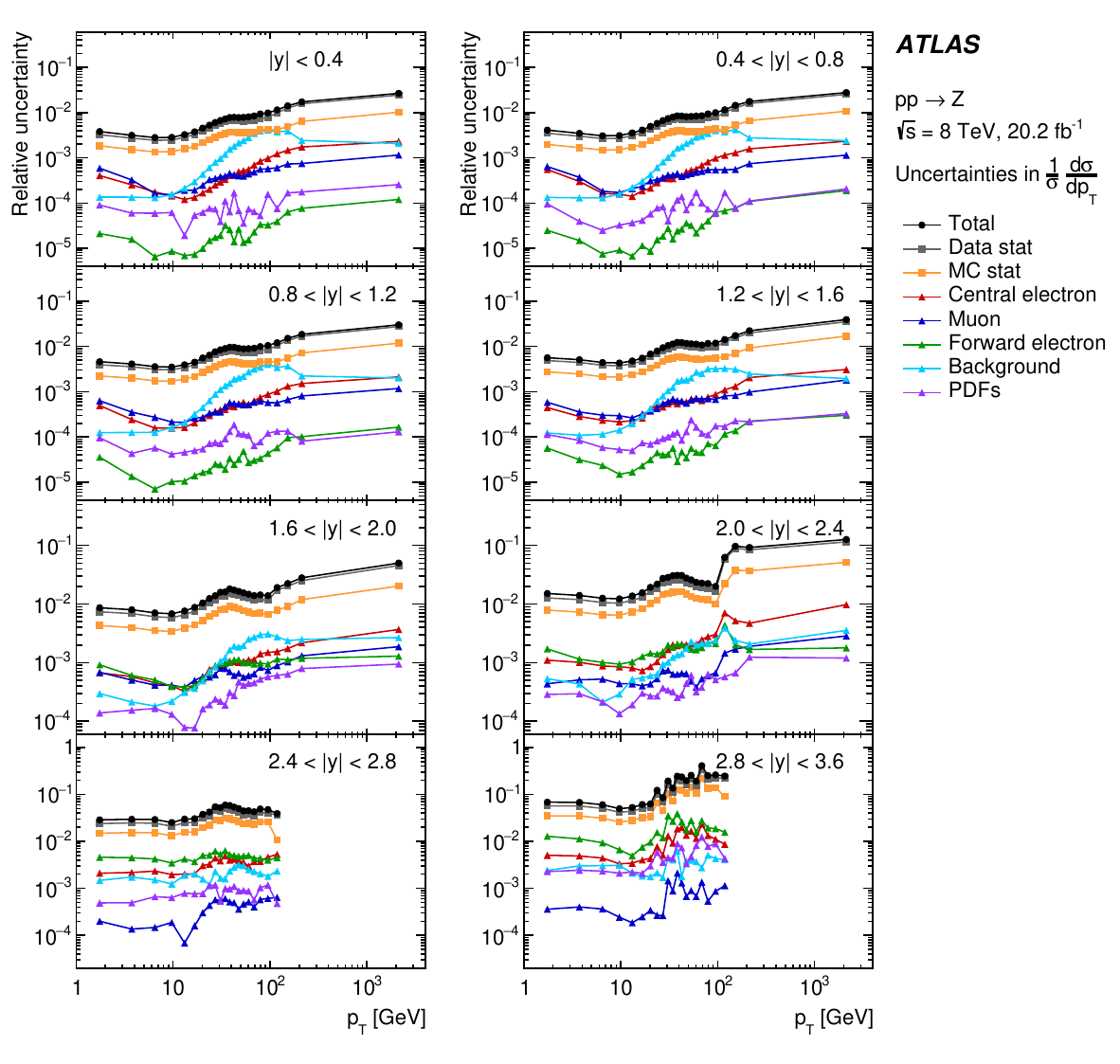}
\caption{
Breakdown of relative uncertainties in the measured normalised differential cross-sections,~$\frac{1}{\sigma}$\dsdpTdy, as a function of~\pTZ. The values are shown, as obtained from the full fit and for each \absyZ~bin, for the combination of the~\eeCC\ and \mmCC~channels for $\absyZ < 1.6$, for the combination of all three channels for $1.6 < \absyZ < 2.4$, and for the \eeCF~channel alone for $2.4 < \absyZ < 3.6$.
\label{uncnorm}}
\end{center}
\end{figure}

This section describes the systematic uncertainties in the measurements of the absolute and normalised differential cross-sections,~\dsdpTdy\ and~\dsdy, as extracted from the angular observables presented in~Fig.~\ref{costheta/phi}. These systematic uncertainties are grouped according to their source, and their typical relative values over most of the kinematic range of the measured observables are listed in~Table~\ref{SysTotX} in the case of the absolute cross-section measurements,~\dsdpTdy. The overall uncertainty of~1.8\% in the integrated luminosity is not shown in this table nor in the figures below, but affects of course all the absolute cross-section measurements presented in this paper. The variations of the uncertainties as a function of~\pTZ\ and~\absyZ\ are shown in~Figs.~\ref{uncpT} (absolute cross-sections) and~~\ref{uncnorm} (normalised cross-sections). The statistical uncertainty in the data is the dominant uncertainty over most of the measurement bins, followed by the statistical uncertainty in the MC~samples. The variations are shown after combining the different channels together, so the uncertainties are correlated through the fit procedure. This explains for example why the forward electron systematic uncertainties do not vanish for~$\absyZ < 1.6$ and why the muon systematic uncertainties do not vanish for~$\absyZ > 2.4$. The breakdown between the different groups of uncertainties is obtained  after freezing all the nuisance parameters of each group to their best-fit values, and then subtracting in quadrature the total uncertainty obtained by performing the fit in these conditions from the total uncertainty obtained by the full fit. Because of the correlations between the nuisance parameters, each of the resulting contributions is rescaled such that their sum in quadrature remains equal to the total uncertainty from the full fit.
 
\begin{figure}[htb!]
\begin{center}
\includegraphics[width=\textwidth]{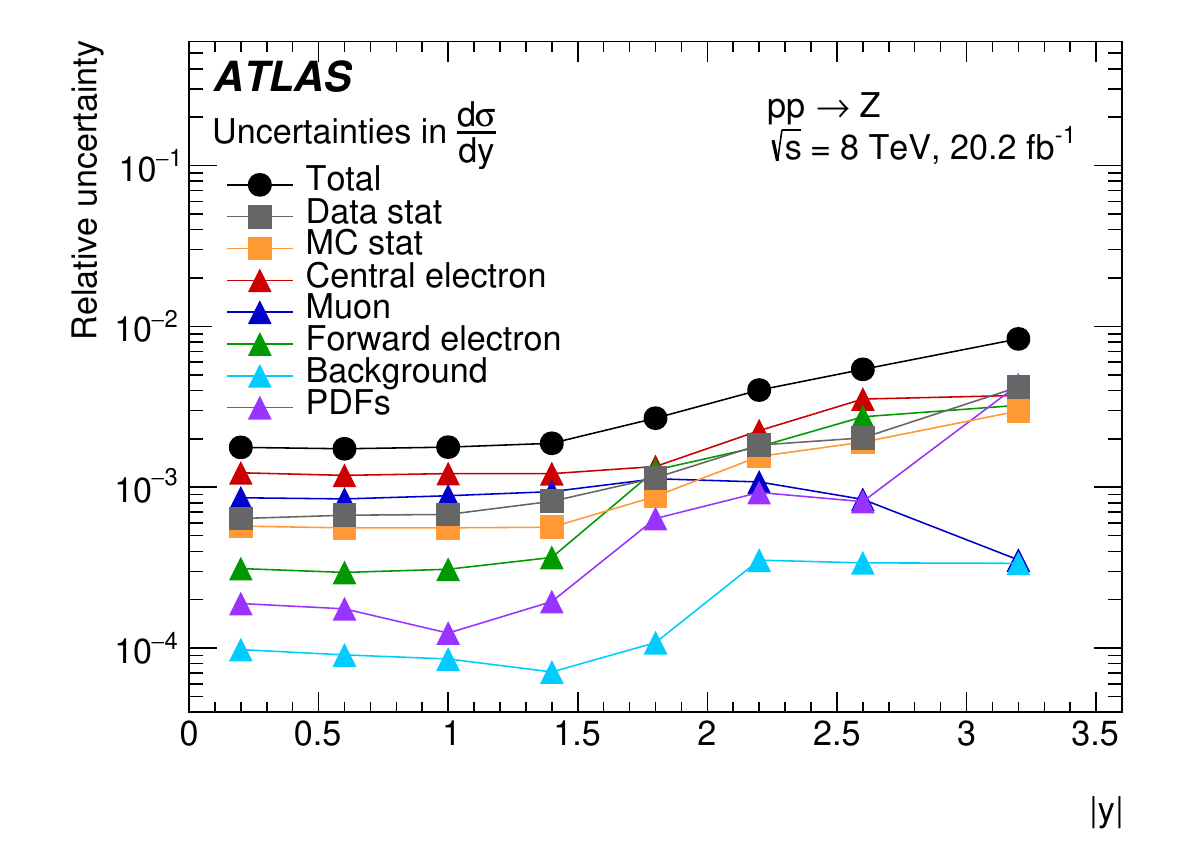}
\caption{
Breakdown of relative uncertainties in the measured absolute differential cross-section,~\dsdy, as a function of~\absyZ. The values are shown, as obtained from the full fit, for the combination of the~\eeCC\ and \mmCC~channels for $\absyZ < 1.6$, for the combination of all three channels for $1.6 < \absyZ < 2.4$, and for the \eeCF~channel alone for $2.4 < \absyZ < 3.6$. The uncertainty of~1.8\% in the integrated luminosity is not included.
\label{uncy}}
\end{center}
\end{figure}

Once these cross-sections are integrated over~\pTZ, the statistical uncertainties obtained for~\dsdy\ are significantly reduced and the experimental systematic uncertainties in the lepton measurements~\cite{PERF-2016-01,PERF-2014-05hack} become dominant, especially in the central region, as shown in~Fig.~\ref{uncy}. In all cases, theoretical uncertainties, arising essentially only from~PDFs, are negligible (see~Ref.~\cite{ATLASAi} for a more detailed discussion). In particular, as mentioned in~Section~\ref{sec:Intro}, higher-order QED/EW effects, such as initial-final state interference diagrammes, which break the factorisation assumption underlying the expression of the differential cross-section in~Eq.~(\ref{eq:master2}), have a negligible impact in the $Z$-boson pole region studied here~\cite{KKMChh19}.

\clearpage
 
\clearpage
 
\begin{figure}[htb!]
\begin{center}
\includegraphics[width=\textwidth]{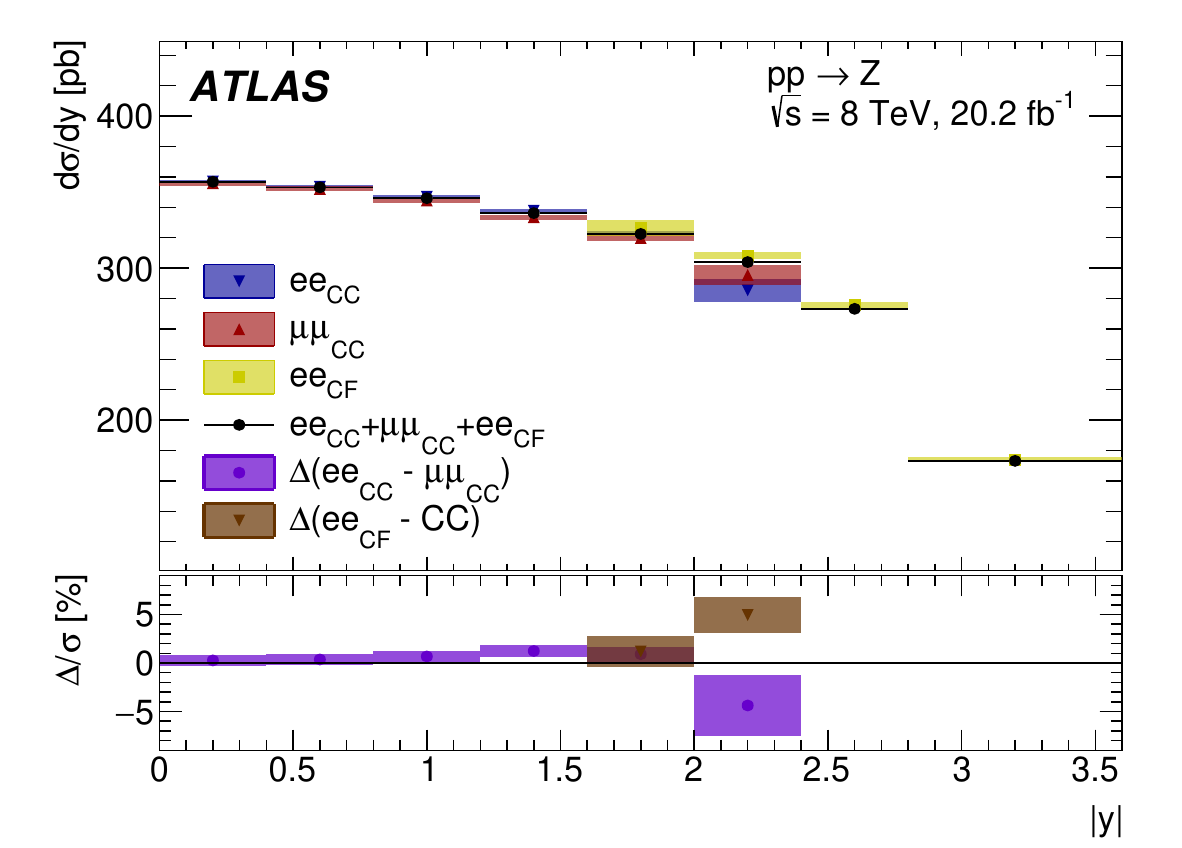}
\vspace{-8mm}
\caption{
Compatibility test between the different channels for the differential \dsdy~cross-section measurements integrated over~\pTZ. The top panel shows the measurements with their total uncertainties (except for the uncertainty of~1.8\% in the integrated luminosity)  for the three channels and their overall combination. The bottom panel shows, for the relevant \absyZ~bins, the relative \dsdy~differences,~$\Delta/\sigma$, in~\%\ between the~\eeCC\ and \mmCC~channels,~$\Delta(\eeCC-\mmCC)$, and between the \eeCF~channel and the combination of the~\eeCC\ and \mmCC~channels,~$\Delta(\eeCF-CC)$.
\label{compatibility}}
\end{center}
\end{figure}
 
\vspace{-8mm}
 
\section{Results and comparisons to predictions}
\label{sec:results}
 
\subsection{Compatibility between measurements}
\label{sec:compatibility}
 
Before considering the measurement results themselves, an important step is to evaluate the compatibility between the different channels. This is first done for~\dsdpTdy\ for the \eeCC~versus \mmCC~channels, which are found to be compatible within their dominant statistical uncertainties in all cases, and are then combined into overall central-central~(CC) measurements. In the results presented in this paper, there are two rapidity bins where the central-central and central-forward measurements overlap, namely in the range~$1.6~<~\absyZ~<~2.4$. The \dsdpTdy~measurements from the \eeCF~channel are compared in this overlap region with those from the CC~channel. The distributions are found to be compatible within their dominant statistical uncertainties, and all three channels are then finally combined. The overall p-value of the combined fit is~4\%, while the overall p-values of fits performed separately for each channel are found to be~11\% for the \eeCC~channel, 90\%~for the \mmCC~channel and 2\%~for the \eeCF~channel.
 
\begin{figure}[htb!]
\begin{center}
\includegraphics[width=\textwidth]{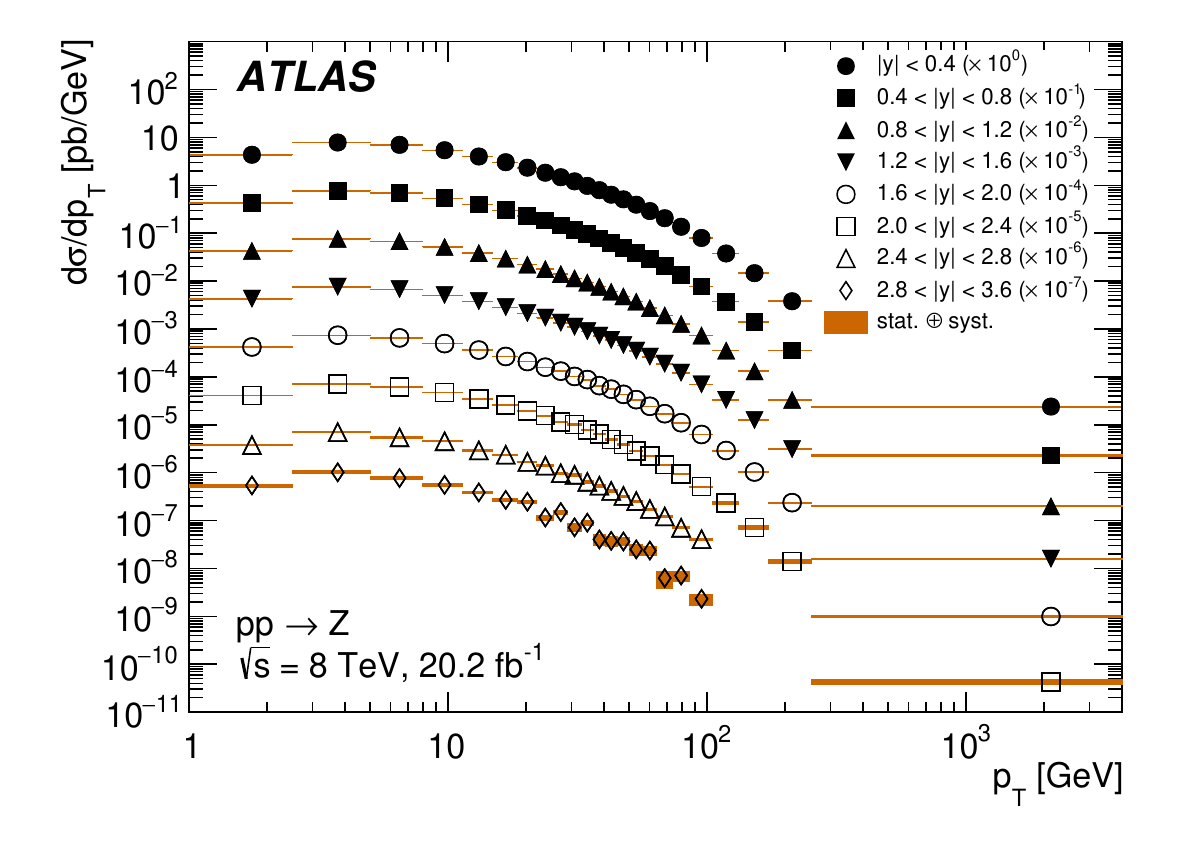}
\caption{
Measured absolute differential \dsdpT\ cross-sections with their total uncertainties shown as a function of~\pTZ\ for each \absyZ~bin. The uncertainty of~1.8\% in the integrated luminosity is not included. For each successive \absyZ~bin, the differential cross section is divided by a factor of ten for plotting purposes.
\label{dsdpTdyspectra}}
\end{center}
\end{figure}
 
A stringent test of the compatibility between channels can be performed for the \dsdy~measurements after integration over~\pTZ, since the statistical uncertainties are strongly reduced as shown in~Fig.~\ref{uncy}. The results of this test are shown as a function of~\absyZ\ in~Fig.~\ref{compatibility}. The top panel of~Fig.~\ref{compatibility} shows comparisons between separate fits done to each of the three channels. The overall p-value for the compatibility between the~\eeCC\ and \mmCC~channels is found to be~2\%, while that between the~\eeCF\ and CC~channels is found to be~3\%. The bottom panel of~Fig.~\ref{compatibility} shows the relative cross-section differences after performing the full combination of all channels between the~\eeCC\ and \mmCC~channels for~$\absyZ < 2.4$ and between the~\eeCF\ and CC~channels for~$1.6 < \absyZ < 2.4$. The residual tensions between channels arise mostly from the highest~\absyZ~bin.
 
\begin{figure}[htb!]
\begin{center}
\includegraphics[width=\textwidth]{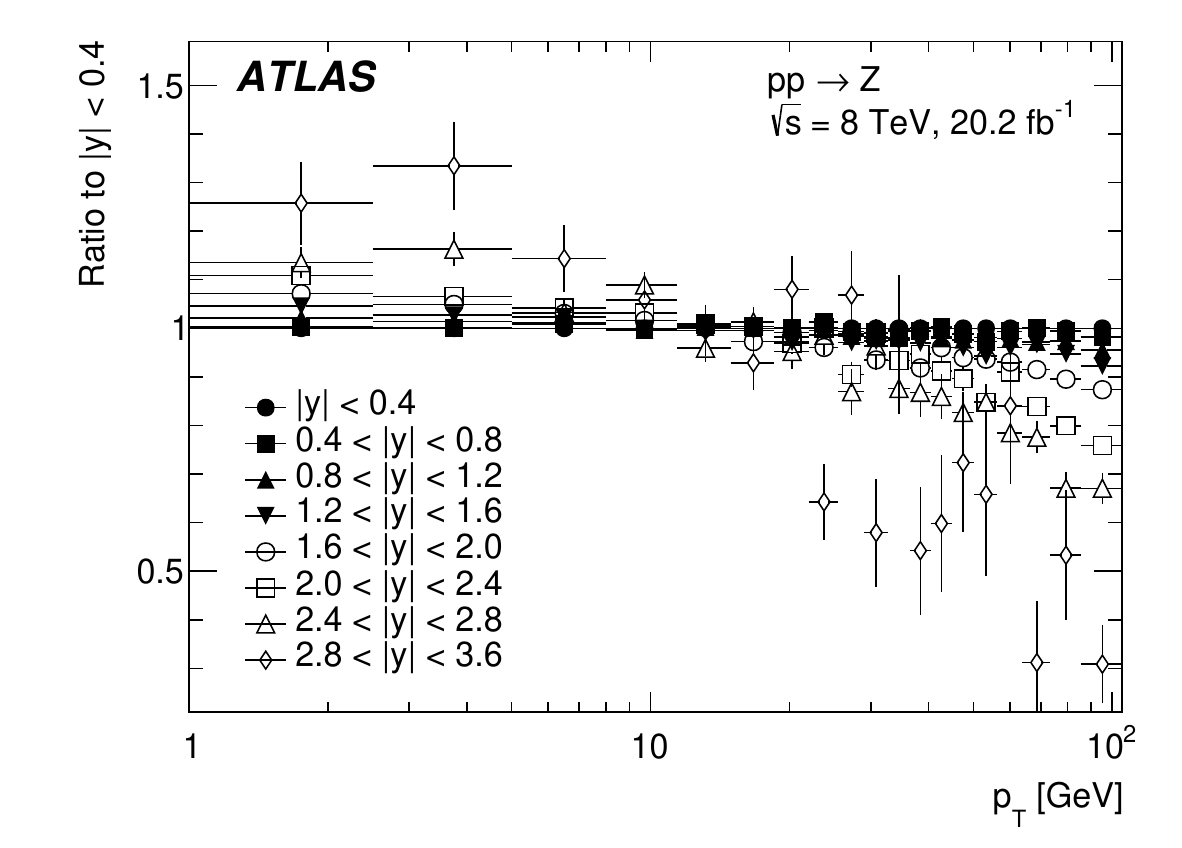}
\caption{
Illustration of the softening of the \pTZ~spectrum at large rapidities. Shown are the ratios of the measured differential \dsdpT~cross-sections in each \absyZ~bin to a reference one taken to be the most central one, $0 < \absyZ < 0.4$. The ratios shown do not include any correlated treatment of the uncertainties between different \yZ~bins.
\label{dsdpTdyspectranorm}}
\end{center}
\end{figure}

\subsection{Comparison between \dsdpTdy~measurements and predictions}
\label{sec:resultsdsdpTdy}
 
For the double-differential \dsdpTdy~measurements, the predictions are obtained by different state-of-the-art QCD perturbative calculations based on \qT~resummation~\cite{Bozzi:2005wk} at approximate approximate~N$^4$LL accuracy. All these calculations, \dyturbo~\cite{Camarda:2019zyx,Camarda:2021ict,Camarda:2023dqn}, CuTe-MCFM~\cite{Neumann:2022lft}, Artemide~\cite{Scimemi:2017etj}, NangaParbat~\cite{Bacchetta:2019sam}, RadISH~\cite{Monni:2016ktx,Bizon:2017rah,Re:2021con}, and SCETlib~\cite{Billis:2021ecs}, are currently being benchmarked in the LHC Standard Model working group. Except for Artemide, they are matched at high values of~\pTZ\ to a common state-of-the-art O($\alpha_s^3$)~fixed-order perturbative calculation from~MCFM~\cite{Neumann:2022lft,Boughezal:2015ded}. Each one of these perturbative resummation calculations has its own recipe for defining a total uncertainty in the prediction based on variations of several QCD~scales related to the resummation and fixed-order contributions and the procedure used to match them such that the summed prediction is well behaved. It is beyond the scope of this paper to explain the differences between the various approaches and the theory predictions are therefore shown below with total uncertainty envelopes provided by the authors.
 
Figure~\ref{dsdpTdyspectra} shows all the absolute differential \dsdpTdy~cross-section measurements with their total uncertainties (except for the uniform uncertainty of~1.8\% in the integrated luminosity). This complete set of cross-section numbers with the full covariance matrix are published in~HEPDATA and contain all the information required for comparisons to theory and for interpretation in terms of PDF~fits or of the strong coupling constant.
As \absyZ~increases, the \pTZ~spectrum becomes softer and this is illustrated in~Fig.~\ref{dsdpTdyspectranorm}, which presents a normalised ratio of the measurements, for which the reference is taken to be the measurement in the most central \absyZ~bin, namely~$0 < \absyZ < 0.4$. Here the luminosity uncertainty cancels, and the softening of the \pTZ~spectrum is clearly visible for the higher \absyZ~bins. This feature, measured for the first time, is expected and well reproduced by the predictions, as shown below. The standard fiducial measurements published for example using the same ATLAS data in~Ref.~\cite{ATLAS8TeV} do not see this kinematic effect because the lepton fiducial selections in~\pT and pseudorapidity largely compensate for it, resulting in a flat dependence of the differential fiducial measurement versus rapidity of the $Z$-boson.
 
\begin{figure}[htb!]
\begin{center}
\includegraphics[width=\textwidth,angle=0]{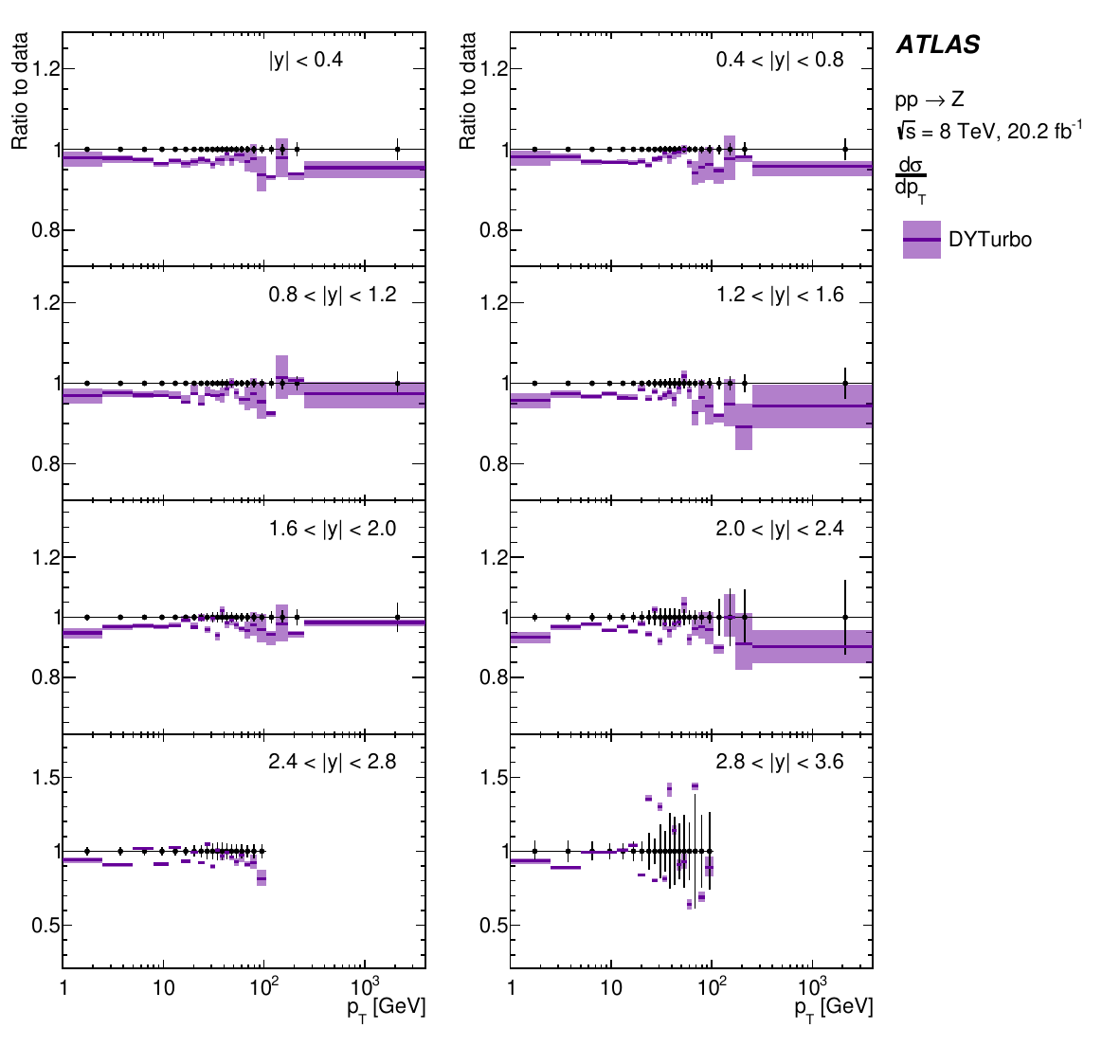}
\caption{
Comparison between the measured absolute differential \dsdpT~cross-sections in each of the eight \absyZ~bins and the predictions from~\dyturbo~\cite{Camarda:2019zyx}. Shown are the ratios between the predictions with their uncertainties as obtained from renormalisation/factorisation/resummation scale variations and the data with their overall uncertainties (except for the uncertainty of~1.8\% in the integrated luminosity). The predictions from~\dyturbo{} are matched to the fixed-order O($\alpha_s^3$) contributions from~MCFM~\cite{Neumann:2022lft,Boughezal:2015ded}.
\label{dsdpTdyDyturbo}}
\end{center}
\end{figure}

\begin{figure}[htb!]
\begin{center}
\includegraphics[width=\textwidth,angle=0]{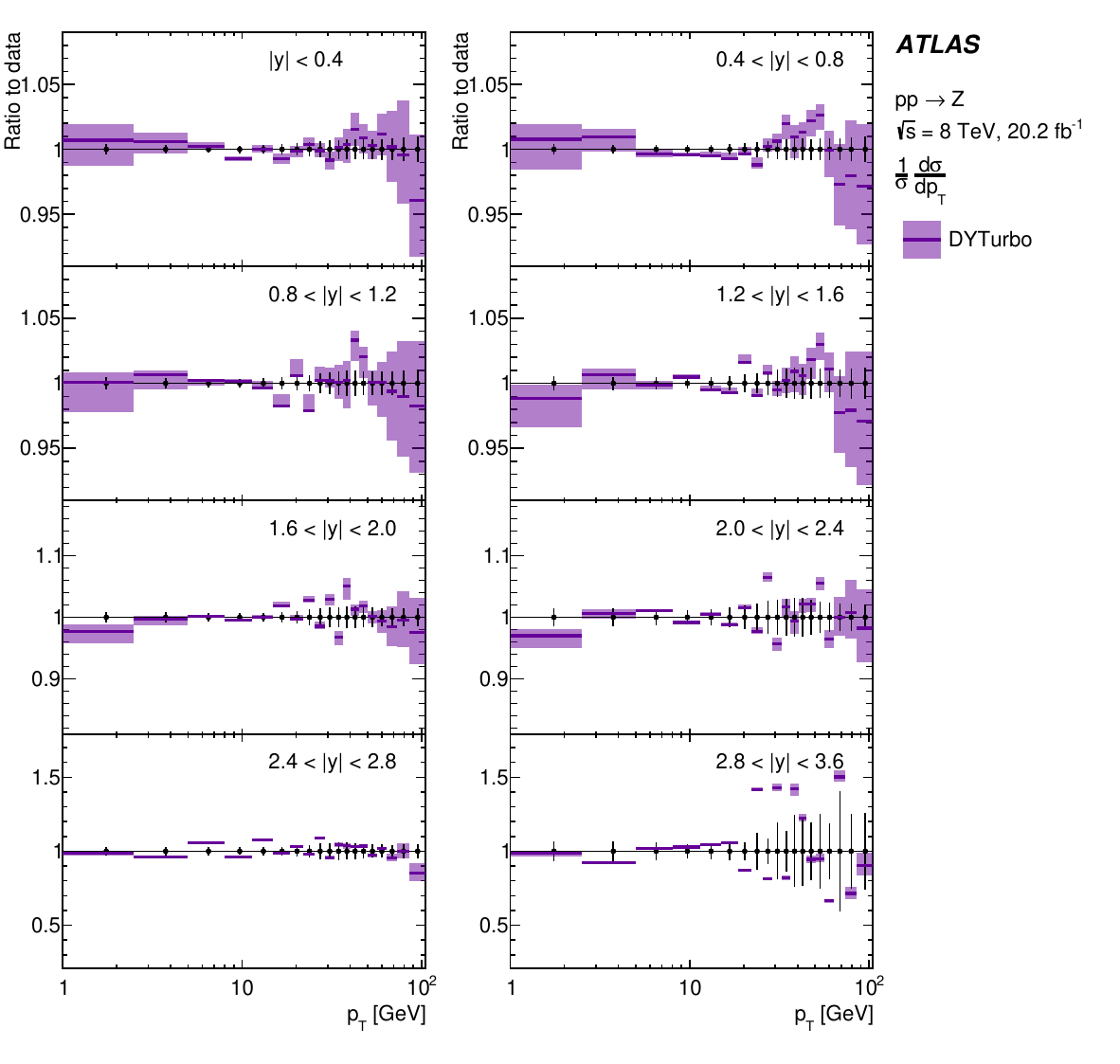}
\caption{
Comparison between the measured normalised differential $\frac{1}{\sigma}$\dsdpT~cross-sections in each of the eight \absyZ~bins and the predictions from~\dyturbo~\cite{Camarda:2019zyx}. Shown are the ratios between the predictions with their uncertainties as obtained from renormalisation/factorisation/resummation scale variations and the data with their overall uncertainties. The predictions from~\dyturbo{} are matched to the fixed-order O($\alpha_s^3$) contributions from~MCFM~\cite{Neumann:2022lft,Boughezal:2015ded}.
\label{dsdpTdyDyturbonorm}}
\end{center}
\end{figure}

Figure~\ref{dsdpTdyDyturbo} shows the ratio of the absolute differential \dsdpTdy~cross-section measurements presented in~Fig.~\ref{dsdpTdyspectra} to the predictions from~\dyturbo. In all \absyZ~bins, the predictions are a few percent lower than the measurements, a feature that is discussed more in detail in~Section~\ref{sec:resultsdsdy}. In the region near the peak of the \pTZ~distribution however, the shape of the predictions is in agreement with that of the measurements within the predominantly theoretical uncertainties. To illustrate this agreement in shape, Fig.~\ref{dsdpTdyDyturbonorm} shows the ratios between the normalised differential measurements and predictions in the range~$0 < \pTZ < 100$~GeV. Over the full \absyZ~range measured, the data and predictions are found to be in agreement within better than~5\%. There are rather strong anti-correlation terms between neighbouring bins, inducing for example the large fluctuations from bin to bin seen for the highest \absyZ~bin at high~\pTZ, where the limited statistics induce terms as large as 40\%.
 
\begin{figure}[htb!]
\begin{center}
\includegraphics[width=\textwidth,angle=0]{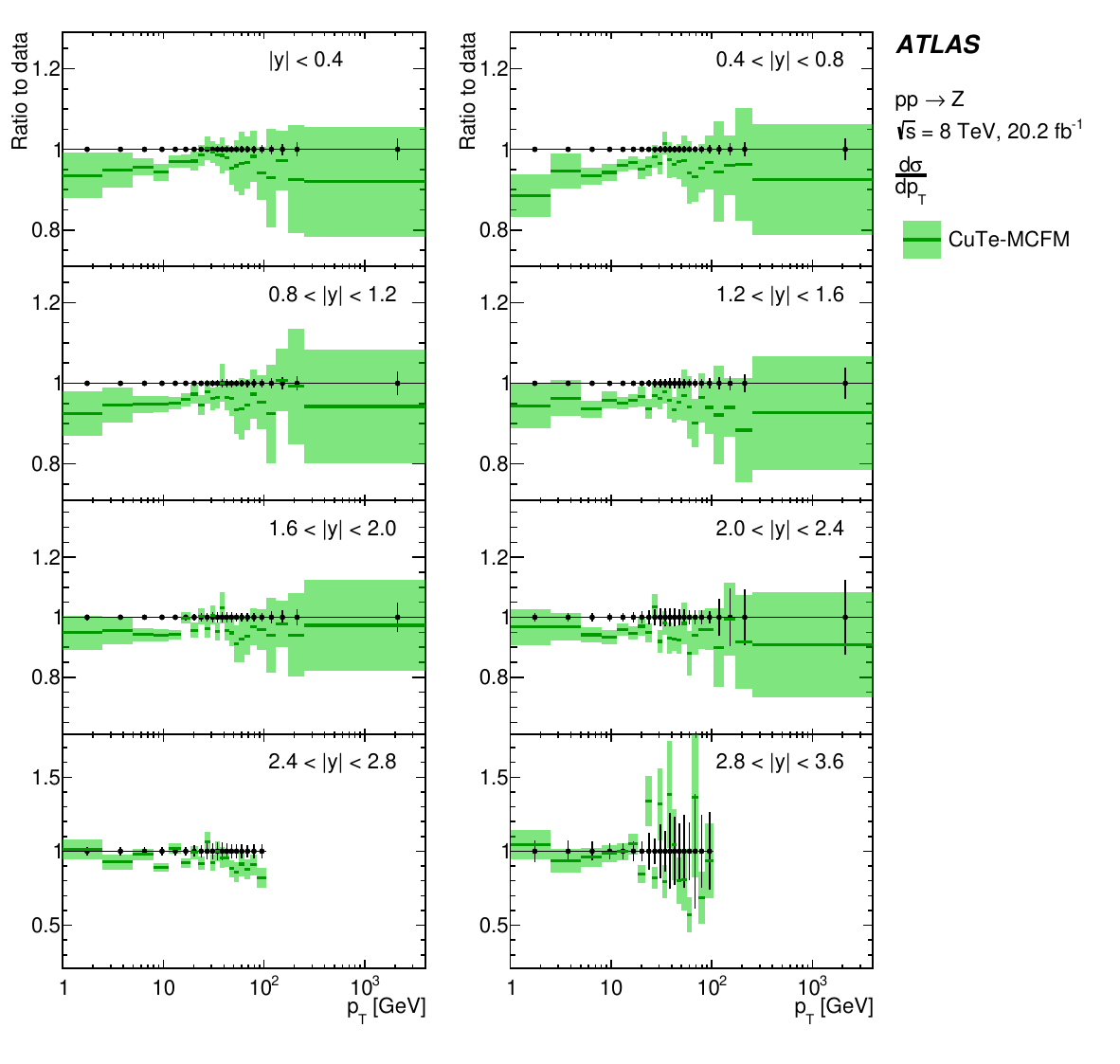}
\caption{
Comparison between the measured absolute differential \dsdpT~cross-sections in each of the eight \absyZ~bins and the predictions from CuTe-MCFM~\cite{Neumann:2022lft}. Shown are the ratios between the predictions with their uncertainties as obtained from renormalisation/factorisation/resummation scale variations and the data with their overall uncertainties (except for the uncertainty of~1.8\% in the integrated luminosity). The predictions from~CuTe-MCFM are matched to the fixed-order O($\alpha_s^3$) contributions from~MCFM~\cite{Neumann:2022lft,Boughezal:2015ded}.
\label{dsdpTdyCute}}
\end{center}
\end{figure}

\begin{figure}[htb!]
\begin{center}
\includegraphics[width=\textwidth,angle=0]{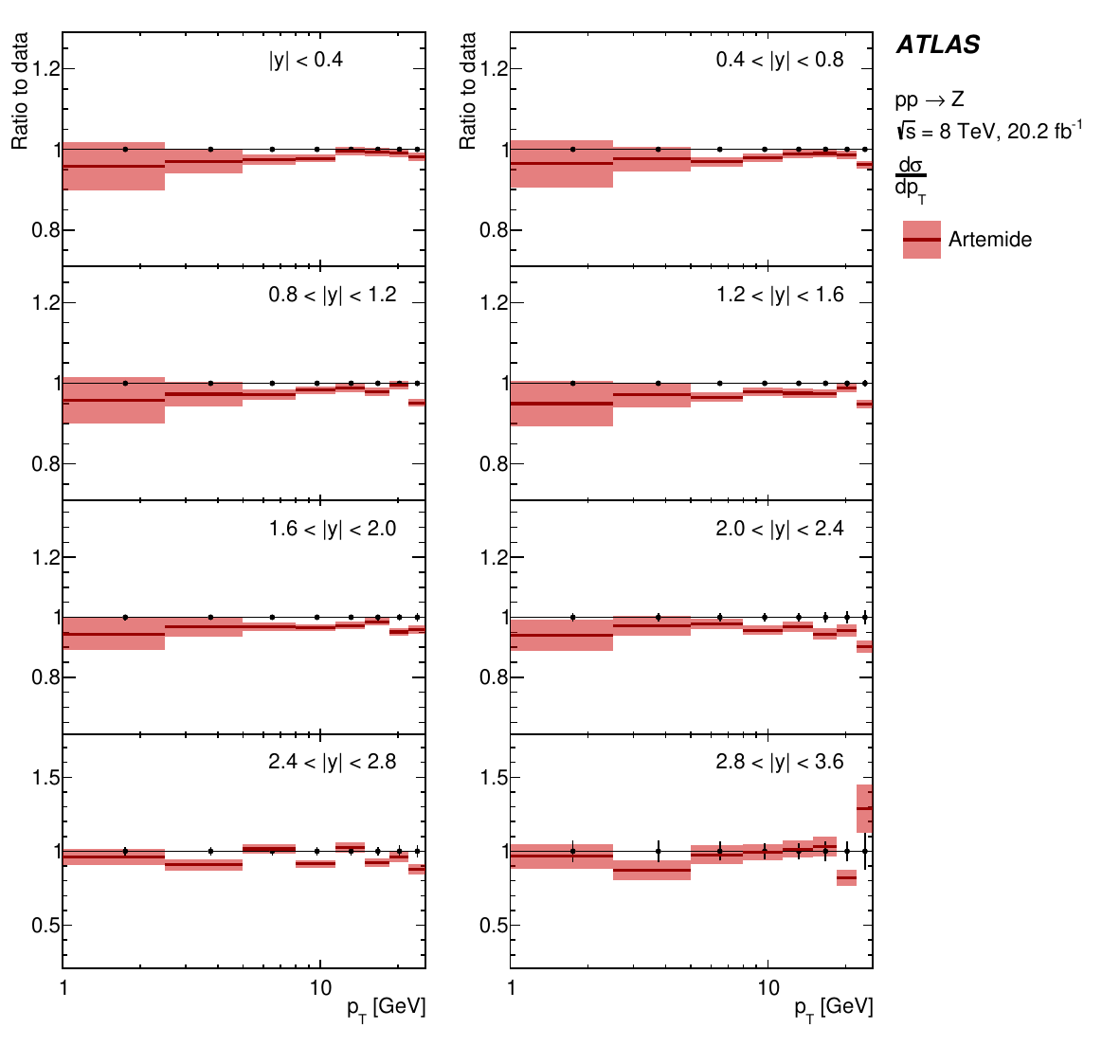}
\caption{
Comparison between the measured absolute differential \dsdpT~cross-sections in each of the eight \absyZ~bins and the predictions from Artemide~\cite{Scimemi:2017etj}. Shown are the ratios between the predictions with their uncertainties as obtained from renormalisation/factorisation/resummation scale variations and the data with their overall uncertainties (except for the uncertainty of~1.8\% in the integrated luminosity). The predictions from~Artemide are from pure resummation and are therefore shown only over the \pTZ~range for which the contribution from fixed-order is below a few~\%. \label{dsdpTdyArtemide}}
\end{center}
\end{figure}

\begin{figure}[htb!]
\begin{center}
\includegraphics[width=\textwidth,angle=0]{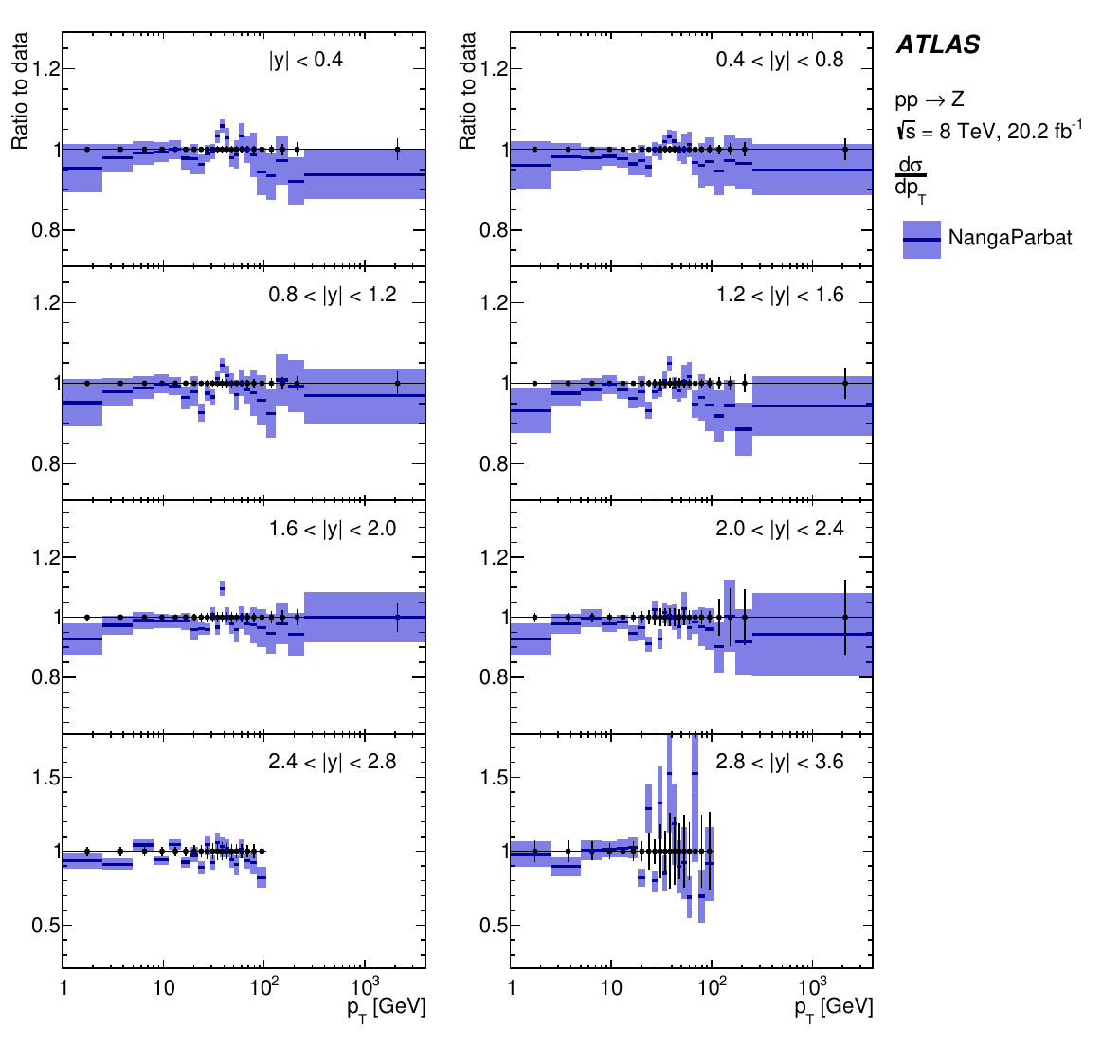}
\caption{
Comparison between the measured absolute differential \dsdpT~cross-sections in each of the eight \absyZ~bins and the predictions from NangaParbat~\cite{Bacchetta:2019sam}. Shown are the ratios between the predictions with their uncertainties as obtained from renormalisation/factorisation/resummation scale variations and the data with their overall uncertainties (except for the uncertainty of~1.8\% in the integrated luminosity). The predictions from~NangaParbat are matched to the fixed-order O($\alpha_s^3$) contributions from~MCFM~\cite{Neumann:2022lft,Boughezal:2015ded}.
\label{dsdpTdyNangaParbat}}
\end{center}
\end{figure}

\begin{figure}[htb!]
\begin{center}
\includegraphics[width=\textwidth,angle=0]{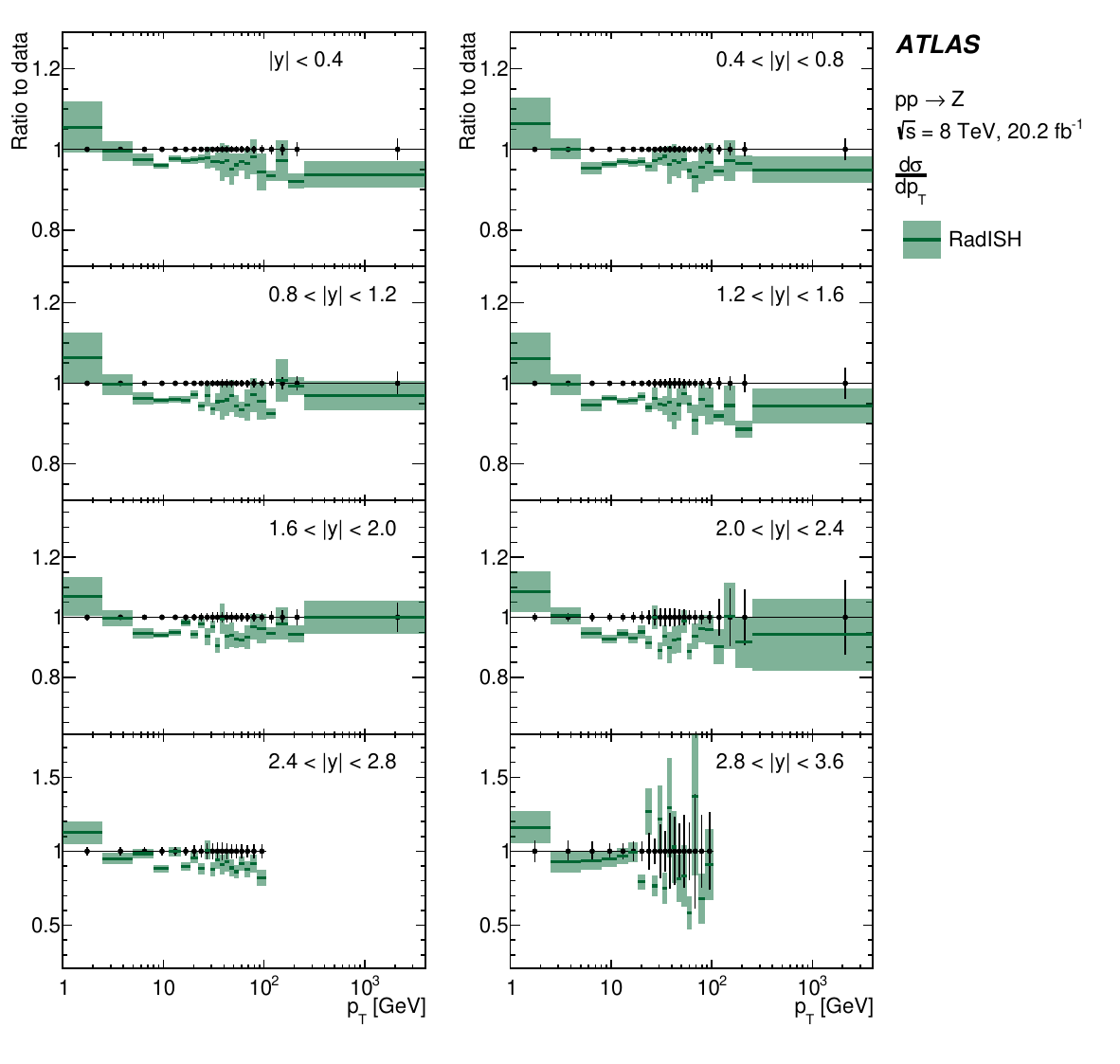}
\caption{
Comparison between the measured absolute differential \dsdpT~cross-sections in each of the eight \absyZ~bins and the predictions from RadISH~\cite{Re:2021con}. Shown are the ratios between the predictions with their uncertainties as obtained from renormalisation/factorisation/resummation scale variations and the data with their overall uncertainties (except for the uncertainty of~1.8\% in the integrated luminosity). The predictions from~RadISH are matched to the fixed-order O($\alpha_s^3$) contributions from~MCFM~\cite{Neumann:2022lft,Boughezal:2015ded}.
\label{dsdpTdyRadish}}
\end{center}
\end{figure}

\begin{figure}[htb!]
\begin{center}
\includegraphics[width=\textwidth,angle=0]{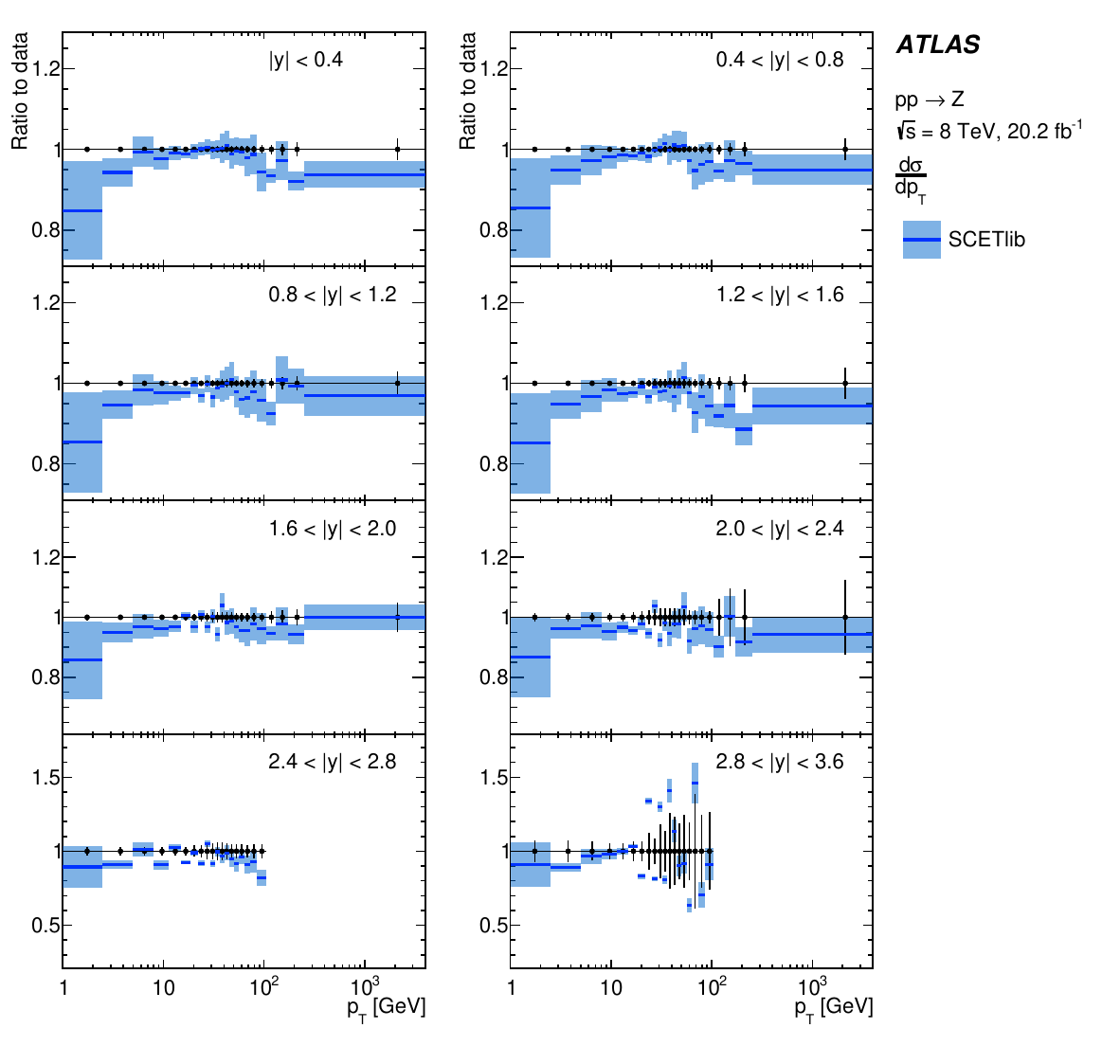}
\caption{
Comparison between the measured absolute differential \dsdpT~cross-sections in each of the eight \absyZ~bins and the predictions from SCETlib~\cite{Billis:2021ecs}. Shown are the ratios between the predictions with their uncertainties as obtained from renormalisation/factorisation/resummation scale variations and the data with their overall uncertainties (except for the uncertainty of~1.8\% in the integrated luminosity). The predictions from~SCETlib are matched to the fixed-order O($\alpha_s^3$) contributions from~MCFM~\cite{Neumann:2022lft,Boughezal:2015ded}.
\label{dsdpTdyScetlib}}
\end{center}
\end{figure}

Figures~\ref{dsdpTdyCute},~\ref{dsdpTdyArtemide},~\ref{dsdpTdyNangaParbat},~\ref{dsdpTdyRadish}, and~\ref{dsdpTdyScetlib} show the same ratios as in~Fig.~\ref{dsdpTdyDyturbo}, successively for the CuTe-MCFM, Artemide, NangaParbat, RadISH, and SCETlib predictions. In all cases, the predictions agree with the measurements within their uncertainties, which range between~2\% and~5\% at the peak of the \pTZ~distribution. The theory uncertainties in all the calculations are arguably incomplete at this stage, since variations between PDF~sets are not included nor are uncertainties at low~\pTZ\ related to heavy quark and non-perturbative effects.
 
\begin{figure}[htb!]
\begin{center}
\includegraphics[width=\textwidth,angle=0]{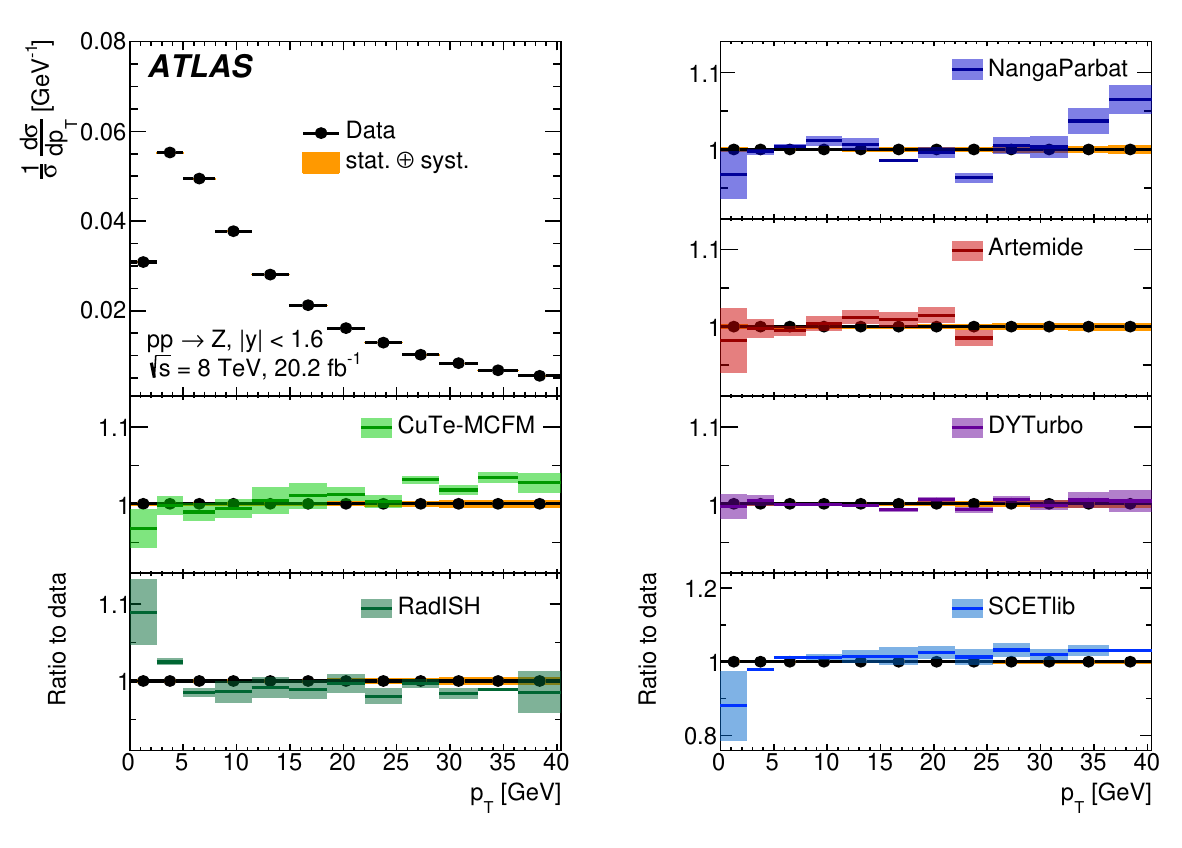}
\caption{
Comparison between the measured normalised differential $\frac{1}{\sigma}$\dsdpT~cross-sections, integrated over~$\absyZ < 1.6$, with their total uncertainties and the predictions from the various resummation calculations. The top left panel shows the data, while the next panels show one by one the ratios between each prediction with its uncertainties as obtained from renormalisation/factorisation/resummation scale variations and the data. Except for Artemide, the predictions are matched to the fixed-order O($\alpha_s^3$) contributions from~MCFM~\cite{Neumann:2022lft,Boughezal:2015ded}.
\label{dsdpTdynormall}}
\end{center}
\end{figure}

To better assess the degree of accuracy with which the different resummation calculations agree with the data, the \dsdpTdy~cross-sections are integrated over~$\absyZ < 1.6$. This improves considerably the statistical uncertainty in the measurements. Figure~\ref{dsdpTdynormall} shows these normalised \absyZ-integrated \dsdpT~spectra for the data and the predictions with their respective uncertainties, together with the ratios between each prediction and the data. The agreement of the predictions with the data is excellent, well within the scale variation uncertainties of~2--3\% over most of the \pTZ~range, except for two points in the lowest \pTZ~bins from the RadISH predictions that are somewhat discrepant.
 
\begin{figure}[htb!]
\begin{center}
\includegraphics[width=\textwidth,angle=0]{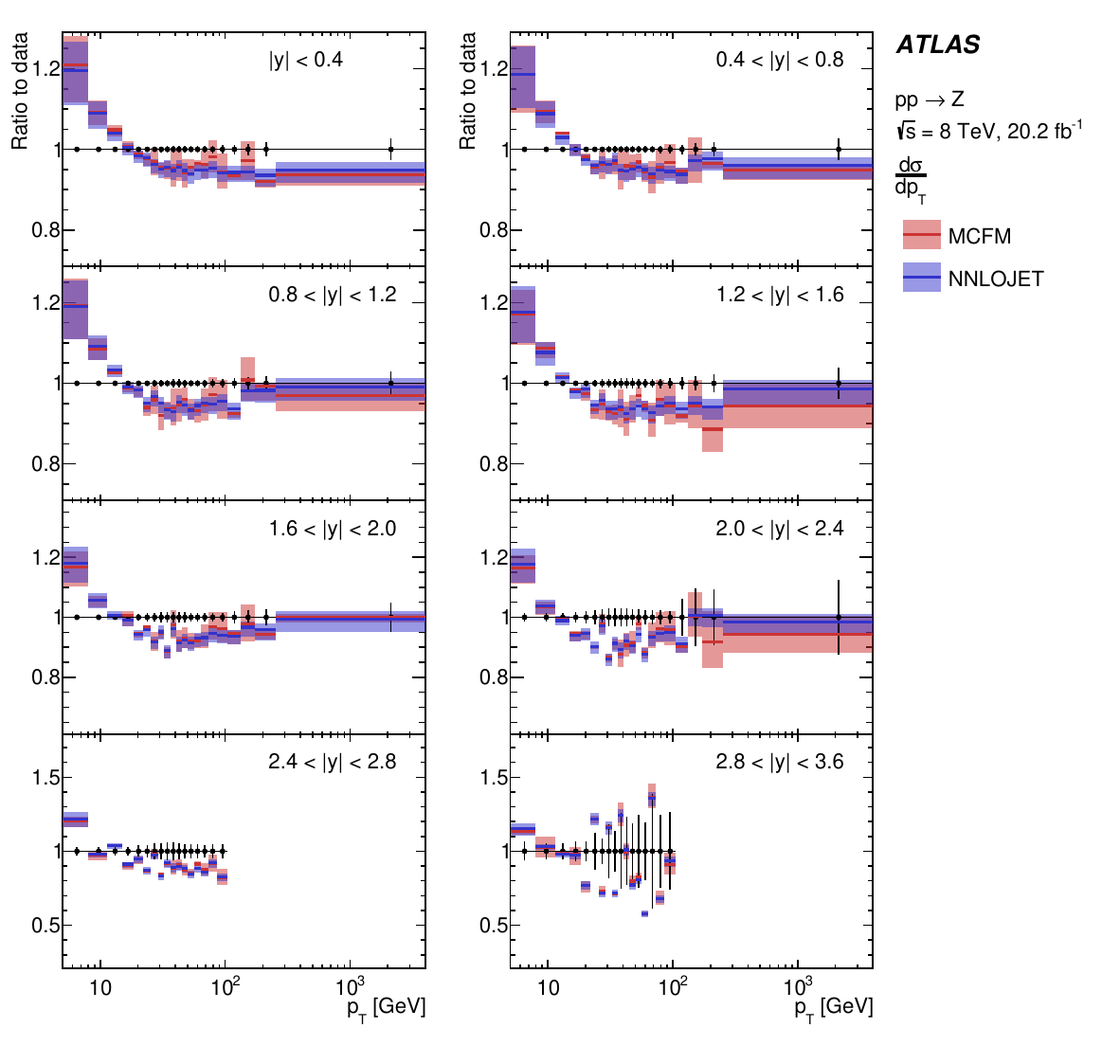}
\caption{
Comparison between the absolute differential \dsdpT~cross-sections with their total uncertainties in each \absyZ~bin and the predictions from the two fixed-order O($\alpha_s^3$) calculations from~MCFM~\cite{Neumann:2022lft,Boughezal:2015ded} and NNLOJET~\cite{Gehrmann_De_Ridder_2016}.
\label{dsdpThighpt}}
\end{center}
\end{figure}

\begin{figure}[htb!]
\begin{center}
\includegraphics[width=\textwidth,angle=0]{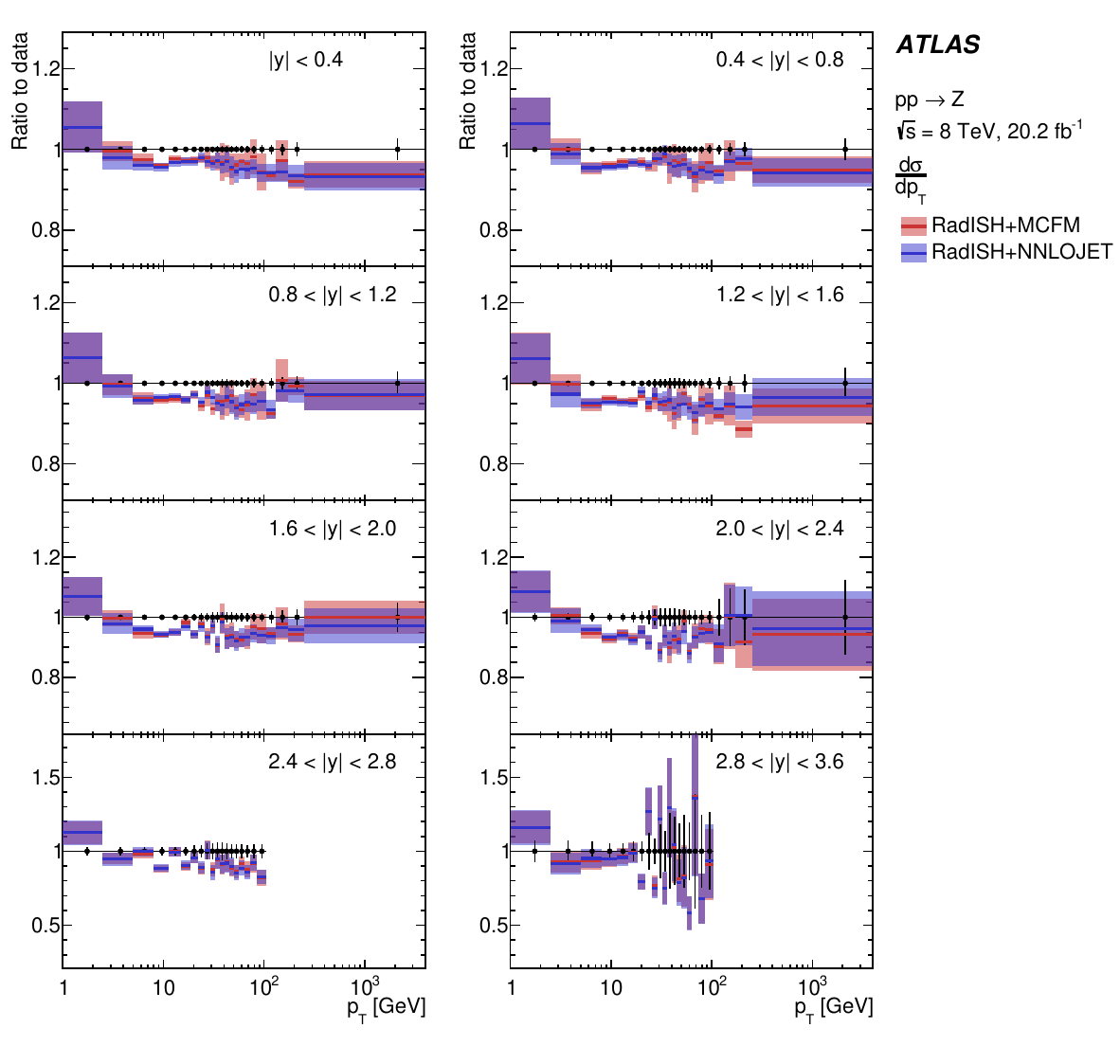}
\caption{
Comparison between the absolute differential \dsdpT~cross-sections with their total uncertainties in each \absyZ~bin and the predictions from RadISH~\cite{Re:2021con} matched to the two fixed-order O($\alpha_s^3$) calculations from~MCFM~\cite{Neumann:2022lft,Boughezal:2015ded} and NNLOJET~\cite{Chen:2022cgv,Gehrmann_De_Ridder_2016}.
\label{dsdpTRadishMCFMvsNNLOJET}}
\end{center}
\end{figure}

Finally, Fig.~\ref{dsdpThighpt} shows a comparison between the measured absolute \dsdpT~differential cross-section in each \absyZ~bin and two fixed-order O($\alpha_s^3$) predictions, that from MCFM, as used by the different calculations discussed above, and that from~NNLOJET~\cite{Gehrmann_De_Ridder_2016}. As expected, these fixed-order calculations are in agreement with the data at high~\pTZ, and do not describe the data well at the lower values of~\pTZ, where resummation is expected to play a major role, as shown for the case of the RadISH predictions matched to each of these fixed-order O($\alpha_s^3$) predictions in~Fig.~\ref{dsdpTRadishMCFMvsNNLOJET}. The larger uncertainties seen in the case of MCFM are mostly due to limited statistics.
 
\clearpage
 
\begin{figure}[htb!]
\begin{center}
\includegraphics[width=\textwidth,angle=0]{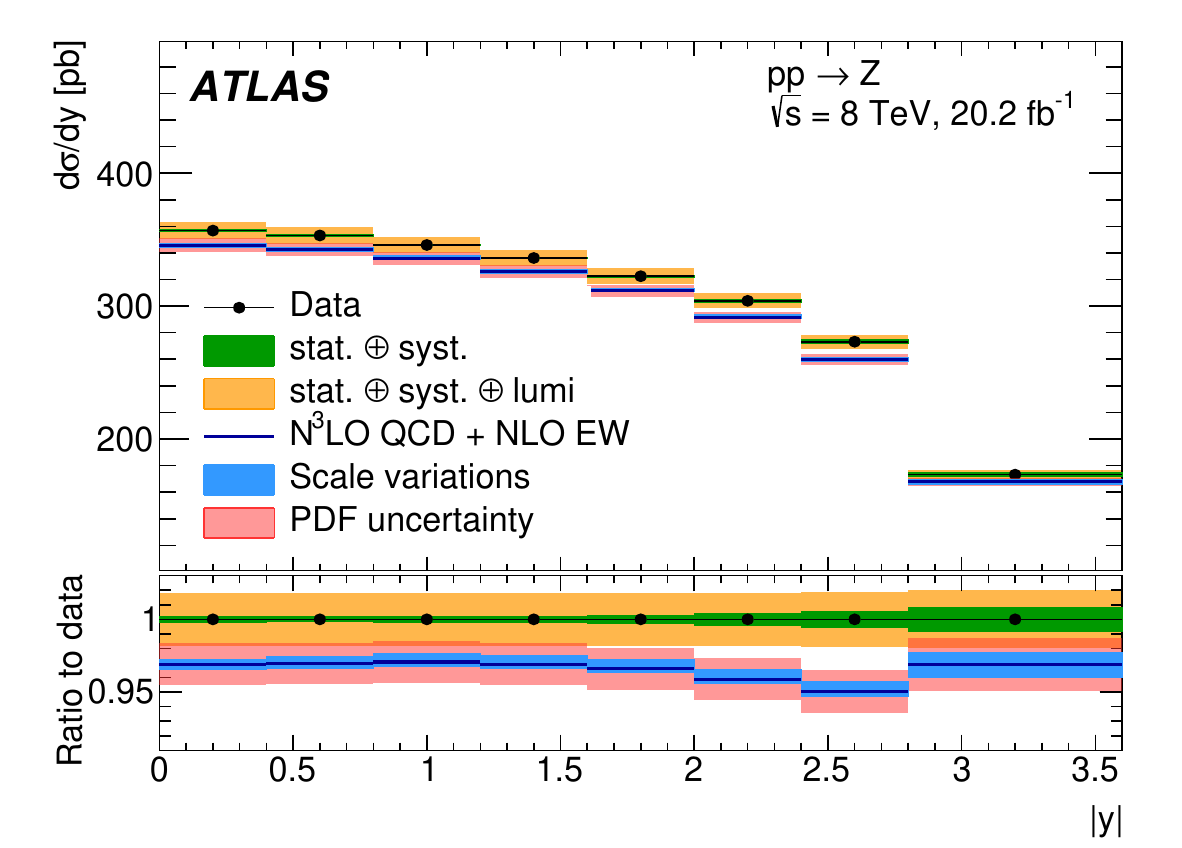}
\caption{Comparison between the \dsdy~measurements and N$^3$LO QCD predictions obtained from~\dyturbo{} using the recent aN$^3$LO~MSHT PDF~set. The N$^3$LO QCD theory prediction includes a negative correction of~0.4\% from NLO EW effects and the theoretical uncertainty bands from QCD renormalisation/factorisation scale variations and from the aN$^3$LO PDF set are shown separately. The bottom panel shows the ratios between predictions and data, where the data uncertainty bands with and without inclusion of the dominant uncertainty of~1.8\% in the integrated luminosity are shown separately.
\label{dsdyN3LO}}
\end{center}
\end{figure}
 
\subsection{Comparison between \dsdy~measurements and predictions}
\label{sec:resultsdsdy}
 
This section focuses on the \dsdy~measurements obtained after integrating the \dsdpTdy~distributions shown above over~\pTZ. As shown in~Fig.~\ref{uncy}, the experimental measurement is significantly more precise with a total uncertainty below~0.2\% in the central rapidity region, owing to the large reduction of the uncertainties of a statistical nature. One also expects the theoretical uncertainties in the predictions to be significantly reduced compared to those shown for the \dsdpTdy~measurements. As a consequence, the focus of the comparisons between data and predictions in this section are on the PDFs, which are expected to provide the dominant uncertainty in the predictions.
 
\begin{table}
\begin{center}
\caption{ \label{dsdyPDFs}
Compatibility test between \dsdy~measurements and predictions obtained from~\dyturbo{} using different PDF~sets. Based on the total uncertainties in the measurements and on the PDF~uncertainties in the predictions shown in~Figs.~\ref{dsdyN3LO}~and~\ref{dsdyNNLO}, the $\chi^2$ per degree of freedom~(d.o.f.) and corresponding p-values are shown for each PDF~set (the uncertainties from CT18A have been rescaled from 95\% to~68\% confidence level). Also shown is the pull on the integrated luminosity in units of its total uncertainty of~1.8\%. }
\begin{tabular}{lccc}
\toprule
PDF set                       & Total $\chi^2$ / d.o.f.  &   $\chi^2$ p-value &     Pull on luminosity \\
\midrule
MSHT20aN$^3$LO~\cite{MSHTN3LO}        & $13/8    $                 &   0.11             &    $ 1.2 \pm 0.6$      \\
CT18A~\cite{Hou_2021}           & $12 / 8  $                 &  0.17              &   $ 0.9 \pm 0.7$       \\
MSHT20~\cite{Bailey_2021}      & $10 / 8  $                 &  0.26              &   $ 0.9 \pm 0.6$       \\
NNPDF4.0~\cite{Ball_2022}      & $30 / 8  $                 &  0.0002            &   $ 0.0 \pm 0.2$       \\
ABMP16~\cite{Alekhin_2017,ABMPupdate}     & $30 / 8  $                 &  0.0002            &   $ 1.8 \pm 0.4$       \\
HERAPDF2.0~\cite{HERAPDF2}     & $22 / 8  $                 &  0.005             &   $ -1.3 \pm 0.8$      \\
ATLASpdf21~\cite{ATLASPDF21}   & $20/8    $                 &  0.01 &   $ -1.1 \pm 0.8$      \\
\bottomrule
\end{tabular}
\end{center}
\vspace{-3mm}
\end{table}

Given the experimental accuracy achieved for the measurements presented in this section, higher-order effects from QED initial-state radiation~(ISR) and from so-called genuine electroweak~(EW) virtual corrections are considered at next-to-leading order and their sum is labelled~NLO~EW. These are directly computed using the code from~Ref.~\cite{Bondarenko:2022mbi}, and are in agreement with earlier results from calculations benchmarked in the LHC EW working group~\cite{PowhegEWs2w,MCSANC,KKMChh19,RADY,Richter-Was:2020jlt}. At the $Z$~pole, the virtual effects decrease the predicted cross-sections by~0.8\%, while the QED~ISR effects increase them by~0.4\%. These corrections are found to be independent of rapidity. All QCD predictions shown below have therefore been decreased by a \absyZ-independent amount of~0.4\% labelled as NLO~EW~corrections.
 
\begin{figure}[htb!]
\begin{center}
\includegraphics[width=\textwidth,angle=0]{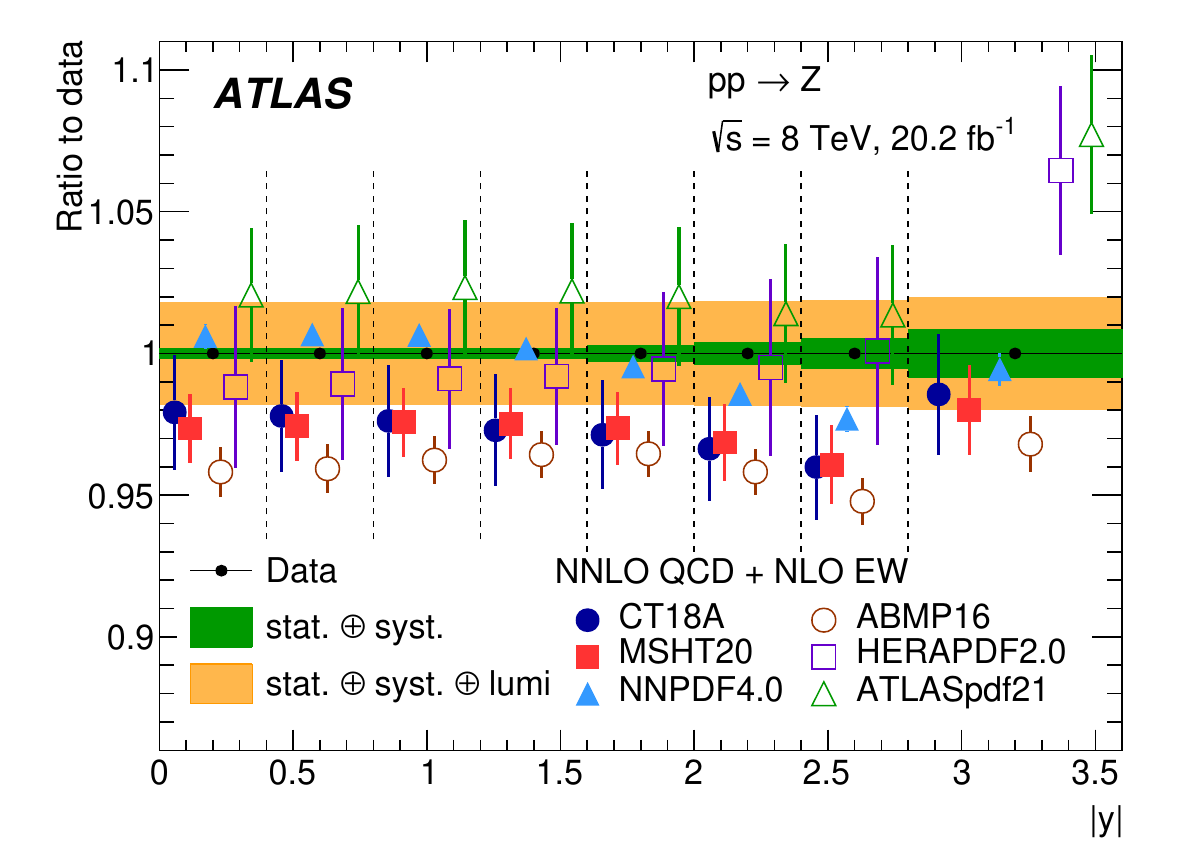}
\caption{Ratio comparison between the \dsdy~measurements and NNLO QCD predictions obtained from~\dyturbo{} using different NNLO PDF~sets. The uncertainty bands in the predictions only show the uncertainties specific to each PDF~set (the uncertainties from CT18A have been rescaled from 95\% to~68\% confidence level). In each \absyZ~bin, the ratios for each PDF~set are gradually displaced to the right for plotting purposes.
\label{dsdyNNLO}}
\end{center}
\end{figure}

In this case, the comparisons to predictions do not depend on \qT~resummation which, through unitarity constraints, does not affect the \pTZ-integrated differential cross-sections. The measurements are thus directly compared to the O($\alpha_s^3$)~fixed-order perturbative predictions from~\dyturbo, supplemented by MCFM~\cite{Neumann:2022lft,Boughezal:2015ded} for the $Z$+jet contribution at O($\alpha_s^3$). These predictions are formally N$^3$LO in~QCD and are obtained using the very recent aN$^3$LO PDF set of Ref.~\cite{MSHTN3LO}. For the first time, such a comparison between experimental measurements and predictions of this formal accuracy is possible and shown in~Figure~\ref{dsdyN3LO}. The theoretical uncertainty bands from QCD renormalisation/factorisation scale variations and from the aN$^3$LO PDF set are shown separately. The uncertainty arising from the scale variations rises slowly from~0.4\% to~1.0\% as \absyZ~increases, while the MSHT PDF uncertainty is constant at around~1.5\%. As shown in the first line of~Table~\ref{dsdyPDFs}, the compatibility of the theory with the data is reasonable, with a p-value of~11\% if one only includes the uncertainties in the~PDFs for the predictions, a standard practice for many publications because most PDF~sets do not usually provide scale variation uncertainties. If one combines the PDF~uncertainties with a combined scale variation uncertainty from~\dyturbo, assumed to be uncorrelated with that from the~PDFs, the p-value obtained is~12\%.

To assess how the aN$^3$LO MSHT20-specific PDF~uncertainty band compares to the presumably non-negligible spread between different PDF~sets, the calculation has been performed one order lower, at~NNLO in~QCD using six of the most recent NNLO~PDF~sets. These comprise CT18A~\cite{Hou_2021}, MSHT20~\cite{Bailey_2021}, NNPDF4.0~\cite{Ball_2022}, ABMP16~\cite{Alekhin_2017,ABMPupdate}, HERAPDF2.0~\cite{HERAPDF2}, and ATLASpdf21~\cite{ATLASPDF21}.
 
Figure~\ref{dsdyNNLO} shows the results of these comparisons as ratios between the predictions and the data. The uncertainties in the theory predictions reflect here only the PDF~specific uncertainties (the uncertainties from CT18A have been rescaled from 95\% to~68\% confidence level). Table~\ref{dsdyPDFs} quantifies the quality of the agreement between the data and the predictions through the total~$\chi^2$ and its corresponding p-value. Also shown is the pull on the integrated luminosity of the experiment for each PDF~set. Only the aN$^3$LO MSHT, NNLO~CT18A and NNLO~MSHT PDF~sets show reasonable agreement with the data, with a positive pull close to one standard deviation on the luminosity, corresponding to predictions approximately~1.6\% lower than the data. The NNPDF4.0 PDF set with its much smaller uncertainties displays poor agreement with the data. This is due to the shape of the predicted distribution since the pull on the integrated luminosity is small. The ABMP16 PDF set is the one that most strongly pulls the integrated luminosity but its poor agreement with the data is also due to its significant difference in shape with respect to the data. The HERAPDF2.0 set and, to a lesser extent, the ATLASpdf21 set also display poor agreement because of a large discrepancy with the data in the highest \absyZ~bin due to the limited set of data used in these fits.

\begin{table}
\begin{center}
\caption{ \label{sigmatot}
Comparison between the measured total cross-section times branching ratio of $Z \to \ell\ell$, $\sigma_Z$, for $Z/\gamma^*$~production with its uncertainty and predictions obtained from~\dyturbo{} using different PDF~sets. The prediction using the MSHTaN$^3$LO~PDF from Ref.~\cite{MSHTN3LO} shows separately the scale variation and PDF~uncertainties. The uncertainties in the NNLO~predictions include only the PDF~uncertainties from each specific PDF~set (the uncertainties from CT18A have been rescaled from 95\% to~68\% confidence level). }
\vspace{5mm}
\begin{tabular}{lcccccccc}
\toprule
&  $\sigma_Z$ (pb)                             \\
Data                          &   $1055 \pm 19$                              \\
\midrule
MSHT20aN$^3$LO~\cite{MSHTN3LO}        &   $ 1023^{+6}_{-4}$ (scale) $\pm 15$ (PDF)   \\
\midrule
CT18A~\cite{Hou_2021}         &   $1028 \pm 19$                              \\
MSHT20~\cite{Bailey_2021}     &   $1027 \pm 13$                              \\
NNPDF4.0~\cite{Ball_2022}     &    $1054 \pm 4 $                             \\
ABMP16~\cite{Alekhin_2017}    &   $1014 \pm 9 $                              \\
HERAPDF2.0~\cite{HERAPDF2}    &    $1058 \pm 25$                             \\
ATLASpdf21~\cite{ATLASPDF21}  &  $1084 \pm 25$                               \\
\bottomrule
\end{tabular}
\end{center}
\end{table}

Finally, the total cross-section times branching ratio of $Z \to \ell\ell$, $\sigma_Z$, for $Z/\gamma^*$~production in the $Z$-boson pole region, $80 < m < 100$~GeV, and within~$\absyZ < 3.6$ is extracted from the integration of the measured differential \dsdy~cross-section:
 
$\sigma_Z = 1055.3 \pm\ 0.7$ (stat.) $\pm\ 2.2$ (syst.) $\pm\ 19.0$ (lumi.) pb
 
Aside from the dominant uncertainty in the integrated luminosity, the overall systematic uncertainty of~0.2\% in this measurement is dominated by experimental lepton efficiency systematic uncertainties and has a negligible contribution from theory uncertainties, which are below~0.1\% and arise essentially from~PDFs. Table~\ref{sigmatot} compares this measurement to the predictions obtained from~\dyturbo{} with the same PDF~sets as those shown in~Table~\ref{dsdyPDFs} for the rapidity-dependent cross-section~\dsdy.

\section{Conclusions}
\label{sec:summary}
 
This paper presents for the first time a double-differential measurement in~(\pTZ,~\absyZ) of absolute and normalised cross-sections at the $Z$~pole within the full phase space of the decay leptons. This is in contrast to the many previous precise unfolded measurements performed in the fiducial phase space of the decay leptons. The measurements in this paper are obtained through a four-dimensional measurement of the lepton angular distributions as a function of~\pTll\ and~\yll\ for a total sample of approximately 15~million $Z$-boson decays measured within the pole region, $80 < \mll < 100$~GeV, and within the range $\absyll < 3.6$. Such a measurement is achieved by extending and improving the methodology already developed and published for the extraction of the $Z$-boson angular coefficients. A profile likelihood fit extracts at the same time these eight angular coefficients and the corresponding unpolarised cross-section as parameters of interest in each measurement bin in (\pTZ,~\absyZ)~space. The uncertainties in these measurements are mostly statistical in nature and the experimental and theoretical systematics are at the few per mille level over most of the range.
 
The \dsdpTdy~measurements are compared to several state-of-the-art QCD perturbative predictions based on \qT-resummation at approximate N$^4$LL accuracy matched to fixed-order O($\alpha_s^3$) calculations at high~\pTZ. The agreement between the data and the predictions is within~5\% over the whole range of the measurements except in kinematic regions with limited statistics.
 
Once integrated over~\pTZ, the rapidity-dependent cross-sections are measured with an overall accuracy of~0.2--0.5\% before accounting for the uncertainty in the integrated luminosity of~1.8\%. They are compared to the predictions from different PDF~sets, which display a varying degree of agreement with the data.
 
The total cross-section times branching ratio of $Z \to \ell\ell$, $\sigma_Z$, for $Z/\gamma^*$~production in the $Z$-boson pole region, $80 < m < 100$~GeV, and within~$\absyZ < 3.6$, is found to be:
 
$\sigma_Z = 1055.3 \pm\ 0.7$ (stat.) $\pm\ 2.2$ (syst.) $\pm\ 19.0$ (lumi.) pb,
 
in agreement with state-of-the-art predictions at N$^3$LO in QCD.

\section*{Acknowledgements}

% The next lines are included from the .//acknowledgements/Acknowledgements.tex input file

We thank CERN for the very successful operation of the LHC, as well as the
support staff from our institutions without whom ATLAS could not be
operated efficiently.
 
We acknowledge the support of
ANPCyT, Argentina;
YerPhI, Armenia;
ARC, Australia;
BMWFW and FWF, Austria;
ANAS, Azerbaijan;
CNPq and FAPESP, Brazil;
NSERC, NRC and CFI, Canada;
CERN;
ANID, Chile;
CAS, MOST and NSFC, China;
Minciencias, Colombia;
MEYS CR, Czech Republic;
DNRF and DNSRC, Denmark;
IN2P3-CNRS and CEA-DRF/IRFU, France;
SRNSFG, Georgia;
BMBF, HGF and MPG, Germany;
GSRI, Greece;
RGC and Hong Kong SAR, China;
ISF and Benoziyo Center, Israel;
INFN, Italy;
MEXT and JSPS, Japan;
CNRST, Morocco;
NWO, Netherlands;
RCN, Norway;
MEiN, Poland;
FCT, Portugal;
MNE/IFA, Romania;
MESTD, Serbia;
MSSR, Slovakia;
ARRS and MIZ\v{S}, Slovenia;
DSI/NRF, South Africa;
MICINN, Spain;
SRC and Wallenberg Foundation, Sweden;
SERI, SNSF and Cantons of Bern and Geneva, Switzerland;
MOST, Taiwan;
TENMAK, T\"urkiye;
STFC, United Kingdom;
DOE and NSF, United States of America.
In addition, individual groups and members have received support from
BCKDF, CANARIE, Compute Canada and CRC, Canada;
PRIMUS 21/SCI/017 and UNCE SCI/013, Czech Republic;
COST, ERC, ERDF, Horizon 2020 and Marie Sk{\l}odowska-Curie Actions, European Union;
Investissements d'Avenir Labex, Investissements d'Avenir Idex and ANR, France;
DFG and AvH Foundation, Germany;
Herakleitos, Thales and Aristeia programmes co-financed by EU-ESF and the Greek NSRF, Greece;
BSF-NSF and MINERVA, Israel;
Norwegian Financial Mechanism 2014-2021, Norway;
NCN and NAWA, Poland;
La Caixa Banking Foundation, CERCA Programme Generalitat de Catalunya and PROMETEO and GenT Programmes Generalitat Valenciana, Spain;
G\"{o}ran Gustafssons Stiftelse, Sweden;
The Royal Society and Leverhulme Trust, United Kingdom.
 
The crucial computing support from all WLCG partners is acknowledged gratefully, in particular from CERN, the ATLAS Tier-1 facilities at TRIUMF (Canada), NDGF (Denmark, Norway, Sweden), CC-IN2P3 (France), KIT/GridKA (Germany), INFN-CNAF (Italy), NL-T1 (Netherlands), PIC (Spain), ASGC (Taiwan), RAL (UK) and BNL (USA), the Tier-2 facilities worldwide and large non-WLCG resource providers. Major contributors of computing resources are listed in Ref.~\cite{ATL-SOFT-PUB-2023-001}.

% End of text imported from the .//acknowledgements/Acknowledgements.tex input file

\clearpage
\printbibliography
 
\clearpage
% ATLAS Collaboration author list
% Reference date of STDM-2018-05 is 2023-02-27
% Author list last updated on date 14-DEC-23
% Data extracted on 14-Dec-2023 for paper reference STDM-2018-05
% at 1:22pm
 
\begin{flushleft}
\hypersetup{urlcolor=black}
{\Large The ATLAS Collaboration}

\bigskip

\AtlasOrcid[0000-0002-6665-4934]{G.~Aad}$^\textrm{\scriptsize 102}$,
\AtlasOrcid[0000-0002-5888-2734]{B.~Abbott}$^\textrm{\scriptsize 120}$,
\AtlasOrcid[0000-0002-1002-1652]{K.~Abeling}$^\textrm{\scriptsize 55}$,
\AtlasOrcid[0000-0001-5763-2760]{N.J.~Abicht}$^\textrm{\scriptsize 49}$,
\AtlasOrcid[0000-0002-8496-9294]{S.H.~Abidi}$^\textrm{\scriptsize 29}$,
\AtlasOrcid[0000-0002-9987-2292]{A.~Aboulhorma}$^\textrm{\scriptsize 35e}$,
\AtlasOrcid[0000-0001-5329-6640]{H.~Abramowicz}$^\textrm{\scriptsize 151}$,
\AtlasOrcid[0000-0002-1599-2896]{H.~Abreu}$^\textrm{\scriptsize 150}$,
\AtlasOrcid[0000-0003-0403-3697]{Y.~Abulaiti}$^\textrm{\scriptsize 117}$,
\AtlasOrcid[0000-0003-0762-7204]{A.C.~Abusleme~Hoffman}$^\textrm{\scriptsize 137a}$
% ...
\bigskip
\\

% ...
$^{29}$Physics Department, Brookhaven National Laboratory, Upton NY; United States of America.\\
% ...
$^{35}$$^{(a)}$Facult\'e des Sciences Ain Chock, R\'eseau Universitaire de Physique des Hautes Energies - Universit\'e Hassan II, Casablanca;$^{(b)}$Facult\'{e} des Sciences, Universit\'{e} Ibn-Tofail, K\'{e}nitra;$^{(c)}$Facult\'e des Sciences Semlalia, Universit\'e Cadi Ayyad, LPHEA-Marrakech;$^{(d)}$LPMR, Facult\'e des Sciences, Universit\'e Mohamed Premier, Oujda;$^{(e)}$Facult\'e des sciences, Universit\'e Mohammed V, Rabat;$^{(f)}$Institute of Applied Physics, Mohammed VI Polytechnic University, Ben Guerir; Morocco.\\
% ...
$^{49}$Fakult\"{a}t Physik , Technische Universit{\"a}t Dortmund, Dortmund; Germany.\\
% ...
$^{55}$II. Physikalisches Institut, Georg-August-Universit\"{a}t G\"ottingen, G\"ottingen; Germany.\\
% ...
$^{102}$CPPM, Aix-Marseille Universit\'e, CNRS/IN2P3, Marseille; France.\\
% ...
$^{117}$Department of Physics, New York University, New York NY; United States of America.\\
% ...
$^{120}$Homer L. Dodge Department of Physics and Astronomy, University of Oklahoma, Norman OK; United States of America.\\
% ...
$^{137}$$^{(a)}$Departamento de F\'isica, Pontificia Universidad Cat\'olica de Chile, Santiago;$^{(b)}$Millennium Institute for Subatomic physics at high energy frontier (SAPHIR), Santiago;$^{(c)}$Instituto de Investigaci\'on Multidisciplinario en Ciencia y Tecnolog\'ia, y Departamento de F\'isica, Universidad de La Serena;$^{(d)}$Universidad Andres Bello, Department of Physics, Santiago;$^{(e)}$Instituto de Alta Investigaci\'on, Universidad de Tarapac\'a, Arica;$^{(f)}$Departamento de F\'isica, Universidad T\'ecnica Federico Santa Mar\'ia, Valpara\'iso; Chile.\\
% ...
$^{150}$Department of Physics, Technion, Israel Institute of Technology, Haifa; Israel.\\
$^{151}$Raymond and Beverly Sackler School of Physics and Astronomy, Tel Aviv University, Tel Aviv; Israel.\\
% ...
$^{a}$ Also Affiliated with an institute covered by a cooperation agreement with CERN.\\
$^{b}$ Also at An-Najah National University, Nablus; Palestine.\\
$^{c}$ Also at Borough of Manhattan Community College, City University of New York, New York NY; United States of America.\\
$^{d}$ Also at Center for High Energy Physics, Peking University; China.\\
$^{e}$ Also at Center for Interdisciplinary Research and Innovation (CIRI-AUTH), Thessaloniki; Greece.\\
$^{f}$ Also at Centro Studi e Ricerche Enrico Fermi; Italy.\\
% ...

\end{flushleft}

% Created with Glance <Atlas.Glance@cern.ch>

\clearpage
 
\end{document}